\documentclass[fleqn,usenatbib]{mnras}

\usepackage{newtxtext,newtxmath}

\usepackage[T1]{fontenc}
\usepackage{longtable}     
\usepackage{bm}      
\usepackage{url}     
\usepackage{graphicx,amsmath,times}     
\usepackage[normalem]{ulem}     
\usepackage{tablefootnote,caption}
\usepackage[dvipsnames]{xcolor}

\DeclareRobustCommand{\VAN}[3]{#2}
\let\VANthebibliography\thebibliography
\def\thebibliography{\DeclareRobustCommand{\VAN}[3]{##3}\VANthebibliography}

\def\simlt{\lower.5ex\hbox{\ltsima}}     
\def\simgt{\lower.5ex\hbox{\gtsima}}     
     
\def\gtsim{\;\lower.6ex\hbox{$\sim$}\kern-6.7pt\raise.4ex\hbox{$>$}\;}     
\def\ltsim{\;\lower.6ex\hbox{$\sim$}\kern-6.9pt\raise.4ex\hbox{$<$}\;}

\def\deg{${}^\circ$}

\definecolor{operamauve}{rgb}{0.718, 0.518, 0.655}

\definecolor{red}{rgb}{1., 0., 0.}

\usepackage{graphicx}	
\usepackage{amsmath}	

\title[RRds in Fornax and in nearby dwarf galaxies]{On the Use of Field RR Lyrae as Galactic Probes. VI. Mixed mode RR Lyrae variables in Fornax and in nearby dwarf galaxies }

\author[V.~F.~Braga et al.]{
V.~F.~Braga $^{1,2,3}$\thanks{E-mail: vittorio.braga@inaf.it}
, G.~Fiorentino $^{1}$
, G.~Bono $^{1,4}$
, P.~B.~Stetson $^{5}$
, C.~E.~Mart{\'i}nez-V{\'a}zquez$^{6}$
, \newauthor{S.~Kwak$^{4}$}
, M.~Tantalo$^{1,4}$
, M.~Dall'Ora $^{7}$
, M.~Di Criscienzo$^{1}$
, M.~Fabrizio $^{1,2}$
, M.~Marengo $^{8}$
, \newauthor{S.~Marinoni $^{1,2}$}
, P.~M.~Marrese $^{1,2}$
, M.~Monelli $^{3}$
and M.~Tantalo$^{1,4}$
\\
$^{1}$INAF-Osservatorio Astronomico di Roma, via Frascati 33, 00040 Monte Porzio Catone, Italy\\
$^{2}$Space Science Data Center, via del Politecnico snc, 00133 Roma, Italy\\
$^{3}$Instituto de Astrof\'isica de Canarias, Calle Via Lactea s/n, E38205 La Laguna, Tenerife, Spain\\
$^{4}$Dipartimento di Fisica, Universit\`a di Roma Tor Vergata, via della Ricerca Scientifica 1, 00133 Roma, Italy\\
$^{5}$Herzberg Astronomy and Astrophysics, National Research Council, 5071 West Saanich Road, Victoria, British Columbia V9E 2E7, Canada\\
$^{6}$Gemini Observatory/NSF's NOIRLab, 670 N. A’ohoku Place, Hilo, HI 96720, USA\\
$^{7}$INAF-Osservatorio Astronomico di Capodimonte, Salita Moiariello 16, 80131 Napoli, Italy\\
$^{8}$Department of Physics and Astronomy, Iowa State University, Ames, IA 50011, USA
}

\pubyear{2022}

\begin{document}
\label{firstpage}
\pagerange{\pageref{firstpage}--\pageref{lastpage}}
\maketitle

\begin{abstract}
We investigate the properties of the mixed-mode (RRd) RR Lyrae (RRL) variables in the Fornax dwarf spheroidal (dSph) galaxy by using $B$- and $V$-band time series collected over twenty-four years. We compare the properties of the RRds in Fornax with those in the Magellanic Clouds and in nearby dSphs, with special focus on Sculptor. We found that the ratio of RRds over the total number of RRLs decreases with metallicity. Typically, dSphs have very few RRds with 0.49$\ltsim P_0 \ltsim $0.53 days, but Fornax fills this period gap in the Petersen diagram (ratio between first overtone over fundamental period versus fundamental period). We also found that the distribution in the Petersen diagram of Fornax RRds is similar to SMC RRds, thus suggesting that their old stars have a similar metallicity distribution.
We introduce the Period-Amplitude RatioS (PARS) diagram, a new pulsation diagnostics independent of distance and reddening. We found that LMC RRds in this plane are distributed along a short- and a long-period sequence that we identified as the metal-rich and the metal-poor component. These two groups are also clearly separated in the Petersen and Bailey (luminosity amplitude versus logarithmic period) diagrams. These circumstantial evidence indicates that the two groups have different evolutionary properties. All the pulsation diagnostics adopted in this investigation suggest that old stellar populations in Fornax and Sculptor dSphs underwent different chemical enrichment histories. Fornax RRds are similar to SMC RRds, while Sculptor RRds are more similar to the metal-rich component of the LMC RRds.
\end{abstract}

\begin{keywords}
 	 
stars: variables: RR Lyrae -- galaxies: dwarf -- (galaxies:) Magellanic Clouds
\end{keywords}



\section{Introduction}\label{sect_intro}
There is solid theoretical \citep[][]{christy66,bono94b} and empirical \citep[][]{cox1983} evidence that the pulsation period of the RR Lyrae stars (RRLs) depends on their physical properties: stellar mass ($\mathcal{M}$), luminosity ($\mathcal{L}$), effective temperature ($T_{eff}$) and chemical composition ($X,Y,Z$, that is, H, He and metal abundances by mass fraction). This dependence is rooted in the so-called fundamental pulsation relation or van Albada \& Baker relation \citep{vanalbada71,marconi15}. Although, pulsation and evolutionary properties mainly depend on $\mathcal{M}$, the measurement and the estimate of $\mathcal{M}$ is a longstanding challenging problem for both single and binary stars \citep[][]{valle2014} and only in the last decade, thanks to the spectroscopic study of binaries \citep{pietrzynski2012} and to asteroseismology \citep[][]{molnar2015,netzel2022}, we are obtaining the first estimates. This limitation applies to radially pulsating variables like RRL themselves, Classical Cepheids (CCs) and type II Cepheids. The reader interested in a more quantitative discussion concerning the most recent mass measurements in radial variables is referred to \citet[][and references therein]{kervella2019a,pilecki2021}.

In this context it is worth mentioning that, with a few exceptions \citep[][and references therein]{kovacs2021}, we still lack a dynamical mass measurement of an old, low-mass star during central helium burning phase (Horizontal Branch, HB), and in particular of RRLs.
The exceptions are a few double-mode RRLs (RRd), i.e., RRLs pulsating simultaneously in the Fundamental (F) and in the First-overtone (FO) mode. Luckily, RRds can play a crucial role in addressing this severe empirical limitation. Double-mode pulsators in general (not only RRLs but also CCs) are distributed along a narrow band of the Petersen Diagram (PD, $P_1/P_0$ versus $P_0$, where $P_0$ and $P_1$ are FU and FO  periods). \citet{popielski2000} discussed in detail on the basis of linear pulsation models the impact that $\mathcal{M}$, $\mathcal{L}$, $T_{eff}$ and $Z$ have on the position of RRds in the PD. They found that the period ratio of RRds mainly depends on $\mathcal{M}$ and $Z$. On the empirical side, the use of the PD as a diagnostic to estimate the mass of a double-mode pulsator dates back to \citet[][8 CCs]{petersen1973} while \citet[][]{cox1980} applied it for the first time to an RRd (AQ Leo). While for CCs the pulsation mass did not agree with evolutionary models---at least until when OPAL opacities \citep[][]{iglesias96} were adopted \citep[][]{moskalik1992,bono2001}---for RRds, the technique provided good results. \citet{bono1996} showed also that, adopting updated opacity tables \citep{iglesias96} and non-linear pulsation models of RRLs \citep{bono94a}, one can also reliably separate the $\mathcal{L}$ levels within the PD.

\citet{bragaglia2001} and, later, \citet{coppola15} showed, based on the RRds hosted in dwarf spheroidal (dSph) galaxies  and in Galactic Globular Clusters (GGCs), that the position of RRds in the PD mainly depends on the metallicity of the stellar system, with more metal-rich RRLs having not only shorter periods, but also lower $P_1/P_0$. The clearest empirical evidence of the dependence on the metallicity is provided by the RRd of the Galactic Bulge \citep[][see their Fig.~5]{soszynski2019}, showing a clump at $P_1/P_0\sim$0.74 and an extended tail down to $P_1/P_0\sim$0.73. These evidence points toward a more metal-rich environment with respect to the Halo, GGCs and dSphs for which the $P_1/P_0$ ranges from 0.742 to 0.747.

One of the key distinctive features of HB stars in nearby dSphs is that they are mostly distributed along the truly horizontal portion of the HB (lack of extreme HB stars) and host sizable samples (hundreds) of RRLs.
Nearby dSphs are, together with the Magellanic Clouds (MCs) and the Bulge, the stellar systems hosting the largest number of RRLs, and in turn, of RRds. The latter are less numerous than  F-mode (RRab) and FO-mode (RRc) RRLs, because the region of the Instability Strip (IS) in which RRLs simultaneously pulsate in two different modes (OR region), is narrower than the regions in which either FU or FO modes alone attain a stable limit cycle \citep{marconi15}.

Owing to its large projected size across the sky \citep[tidal radius r $\sim$ 71 arcmin][]{mateo98a}, a full census of Fornax variable star content is still missing. Several attempts have been devoted to characterize some specific properties of the Fornax variable stars. The RRLs have been mainly discussed by \citet{bersier02} (515 RRLs), \citet{greco07,greco2009} (27 and 30 RRLs in two Globular Clusters) and \citet{fiorentino2017b} ($>$1400). More recently, a large catalogue of variable stars---including Fornax---based on Dark Energy Survey (DES) Collaboration data collected during six year with the Dark Energy Camera \citep[DECam][]{stringer2021} was published. However, none of the quoted works was focused on a detailed study of RRds.

In this paper, by using the same dataset adopted in \citet{fiorentino2017b}, we provide pulsation properties for RRds in the Fornax dSph. Their properties have been compared with those found in other dSphs (Sculptor, Carina, Draco, Sagittarius), in the MCs and in the Galactic bulge.
The structure of the paper is the following. In \S~2 we discuss the adopted multi-band optical photometry, while \S~3 deals with the identification of RRds both in Fornax and in Sculptor dSphs. The diagnostics adopted  to investigate pulsation and evolutionary properties of RRds, are discussed in \S~4 together with the comparison among the different RRd samples available in the literature. In \S5 we summarize the results of this investigation and we outline the future developments of this project. 



\section{Optical photometry}\label{sect_data}

\begin{figure}
	\includegraphics[width=\columnwidth]{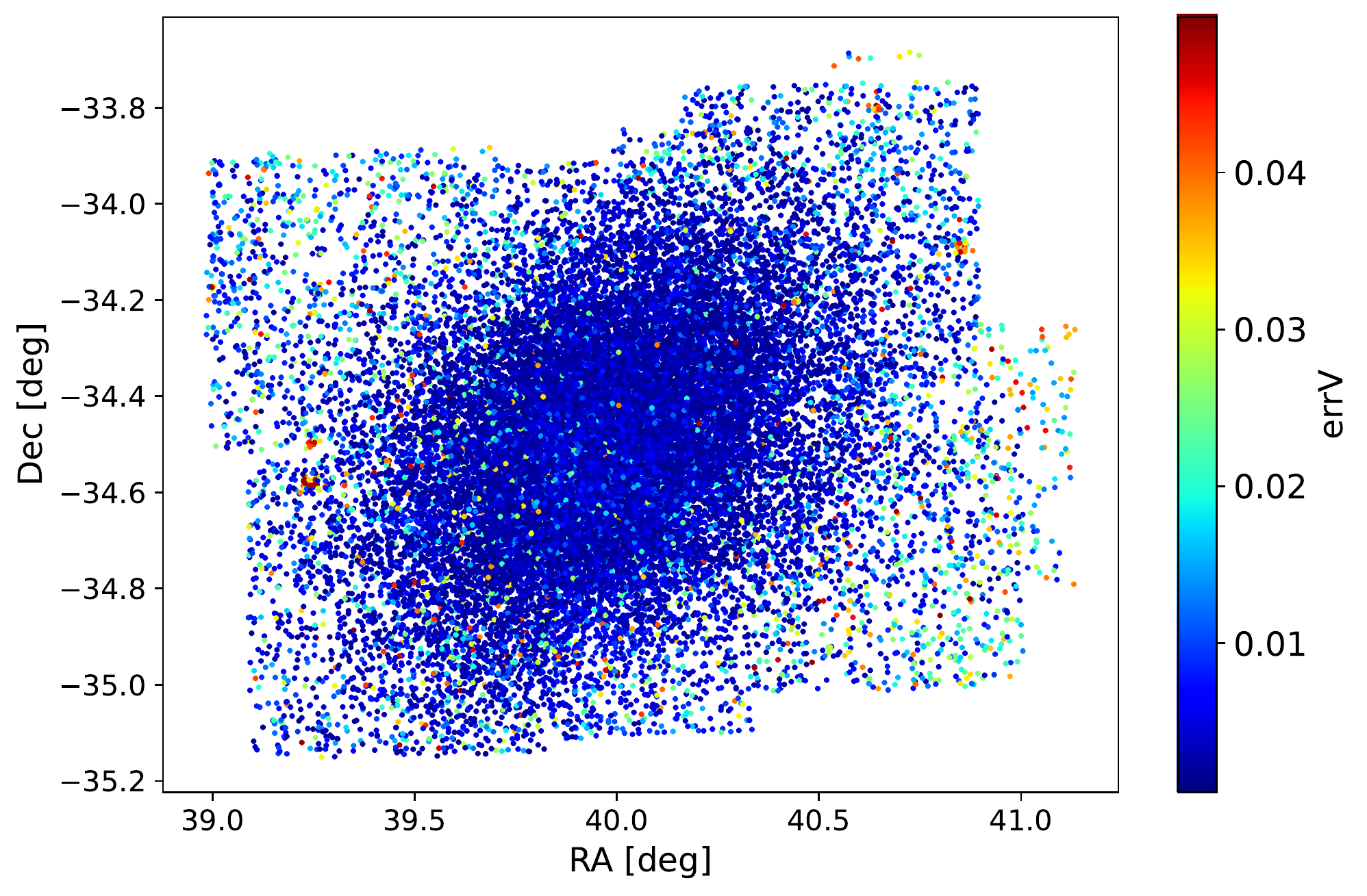}
    \caption{Sky distribution in RA and DEC of all the stars with $V$< 22.5 mag detected by our photometric reduction. The uncertainty on the $V$-band magnitude (repeatability error, $errV$, see \citealt{stetson87}) is color-coded (see the bar on the right).}
    \label{fig:errmag}
\end{figure}

Based on more than 10,000 optical CCD images, collected with ground-based telescopes ranging from 1m- to 8m- during almost two dozen years, we obtained new accurate and homogeneously calibrated {\it BVRI\/} photometry for the stellar populations in the  Fornax dSph. This work is part of a larger project, led by P. B. Stetson, devoted in building up homogeneous photometry for all the Milky Way satellites, including dSphs and GGCs. Our calibrated photometry covers a 2\deg$\times$1.1\deg~ sky area around the galaxy center. The astrometric and photometric measurements were obtained by using well established techniques using the DAOPHOT/ALLSTAR/ALLFRAME suite (see, e.g., \citealp{stetson00,stetson05a}, and references therein). The current photometric catalog includes calibrated photometry for 737,959 stars, with at least one measurement in two different bands, in the field of Fornax. Despite the crowding, even the innermost regions of the galaxy are not affected by blending. Fig.~\ref{fig:errmag} displays the distribution on the sky of all the stars in our photometric catalog down to V$<$22.5 mag. This limiting magnitude  is $\sim$1 mag fainter than the Horizontal Branch and data plotted in this figure show that more than 90\% of the selected stars have an intrinsic error (color-coded) in the $V$-band smaller than 1\%.
The adopted data reduction strategy provides time series in four bands ($BVRI$) and the number of phase points per star in $B$ plus $V$-band ranges from $\sim$30 to $\sim$600, while the number of phase points in $R$ and $I$-band is on average one order of magnitude smaller. The RRd analyzed in this investigation have a number of phase points ranging from $\sim$40 to $\sim$550, in particular, $\sim$90\% of the current RRds have more than 120 $B$ plus $V$ phase points. More details on the data reduction strategy and on spatial coverage of the different data sets will be provided in a forthcoming paper (Braga et al. 2022, in preparation, [B22]).

To compare our dataset to that in Sculptor dSph, we have refined the classification and the pulsation properties of RRds in this galaxy by using the optical database by \citet[][]{martinezvazquez16b}. The two datasets are similar concerning the time coverage (two dozen years) and the data reduction strategy. We have obtained $B$- and $V$-band time series with a number of phase points ranging from more than 40 to $\sim$190, that are well-suited to investigate the periodicity of RRds.

\section{Identification of variable stars in Fornax}\label{sect_ident}

To overcome possible observational biases in defining our list of variable stars and before cross-correlating Fornax external catalogs, we performed a blind variability search. This was done by using five different implementations of period search algorithms: 
i) the Welch-Stetson \citep[WS,][]{welch93,stetson1996} 
index; 
ii) The Lomb-Scargle Periodogram \citep[LS,][]{scargle82}; 
iii) the Generalized Lomb-Scargle Periodogram \citep[GLS,][]{zechmeister2009};  
iv) The GATSPY package based on the Lomb-Scargle Periodogram \citep[][]{vanderplasivezic2015} and
v) the Phase-Dispersion Minimization \citep[PDM,][]{stellingwerf78}. 
We have visually inspected the light curves folded with the best-period estimates provided by the quoted algorithms applied on both the \textit{B}- and \textit{V}-band time series and selected the best one for each candidate variable.

More specifically, we obtained a set of 15 period estimates for each RRL, of which 3($B$)+3($V$) from the LS periodogram, 3($B$)+3($V$) from the GLS periodogram, 1($B$)+1($V$) from the PDM and one from GATSPY. Note that GATSPY performs a simultaneous analysis of $B$- and $V$-band data and provides a single estimate of the period. The periods estimates were ranked according to $\chi$-squared of the folded light curves over the sinusoidal model generated by the period-search algorithms. Subsequently, we visually inspected one-by-one the folded $B$- and $V$-band light curves and checked whether the highest-ranked period was also the best one, i.e., the one showing the lowest dispersion. The visual inspection was requested because both RRab and RRd variables do not have sinusoidal light curves and a blind $\chi$-squared minimization might lead to inaccurate measurements of the period.
Note that the quoted periodicity diagnostics---the different LS algorithms and the PDM---do not provide an estimate of the uncertainty of the period and the bootstraping techniques are computationally prohibitive \citep{vanderplas2018}. Therefore, we adopted frequency grids for both LS and PDM analysis with steps of $\sim$2-5$\cdot 10^{-4} d^{-1}$, i.e. steps in time shorter than $10^{-4}$ days. However, there is some room to subjectivity. In several cases we have to select, among the 15 period estimates, three or four periods that were identical within $10^{-5}$ days. Note that primary and secondary RRd periods were subsequently refined by running the periodicity search algorithm with a frequency grid two times finer. Therefore, we can assume that the relative uncertainty of our periods is on average better than 5$\cdot 10^{-4}$ for all the variables and a factor of two smaller for RRds. More quantitative details concerning the periodicity search algorithms will be provided in B22.

We ended up with a list of 2,266 periodic variables within the field of Fornax, of which 22 were labelled as Cepheids of generic type\footnote{We did not focus our attention on the separation among the different types of Cepheids, because this is not the aim of the current investigation.}, 2,068 RRLs, 91 SX Phoenicis (SXP) and 85 variables for which the classification is uncertain.

\subsection{RRd identification in Fornax}\label{sec:fornax}

We have flagged as candidate multiperiodic RRLs those showing not only a clear periodic behavior but also a dispersion in their light curves folded at the period of the dominant mode ($P_{dom}$). In principle, these stars include both Blazhko RRLs and RRds. To separate these two variable types, we followed both a qualitative and a quantitative approach. 

Based on a visual inspection, we have labelled as Blazhko RRLs the variables only displaying clear amplitude modulations. These are easily spotted because they are characterized by a larger dispersion of the phase points located at maximum and minimum light. A posteriori, we have also checked the unfolded time series data and verified that the change in the shape of the day-to-day light curve was minimal, as expected for Blazhko RRLs. Indeed, their secondary modulation periods are significantly longer than the pulsation period and range from a few days to a few  years \citep[][]{skarka2020}.

\begin{figure}
	\includegraphics[width=\columnwidth]{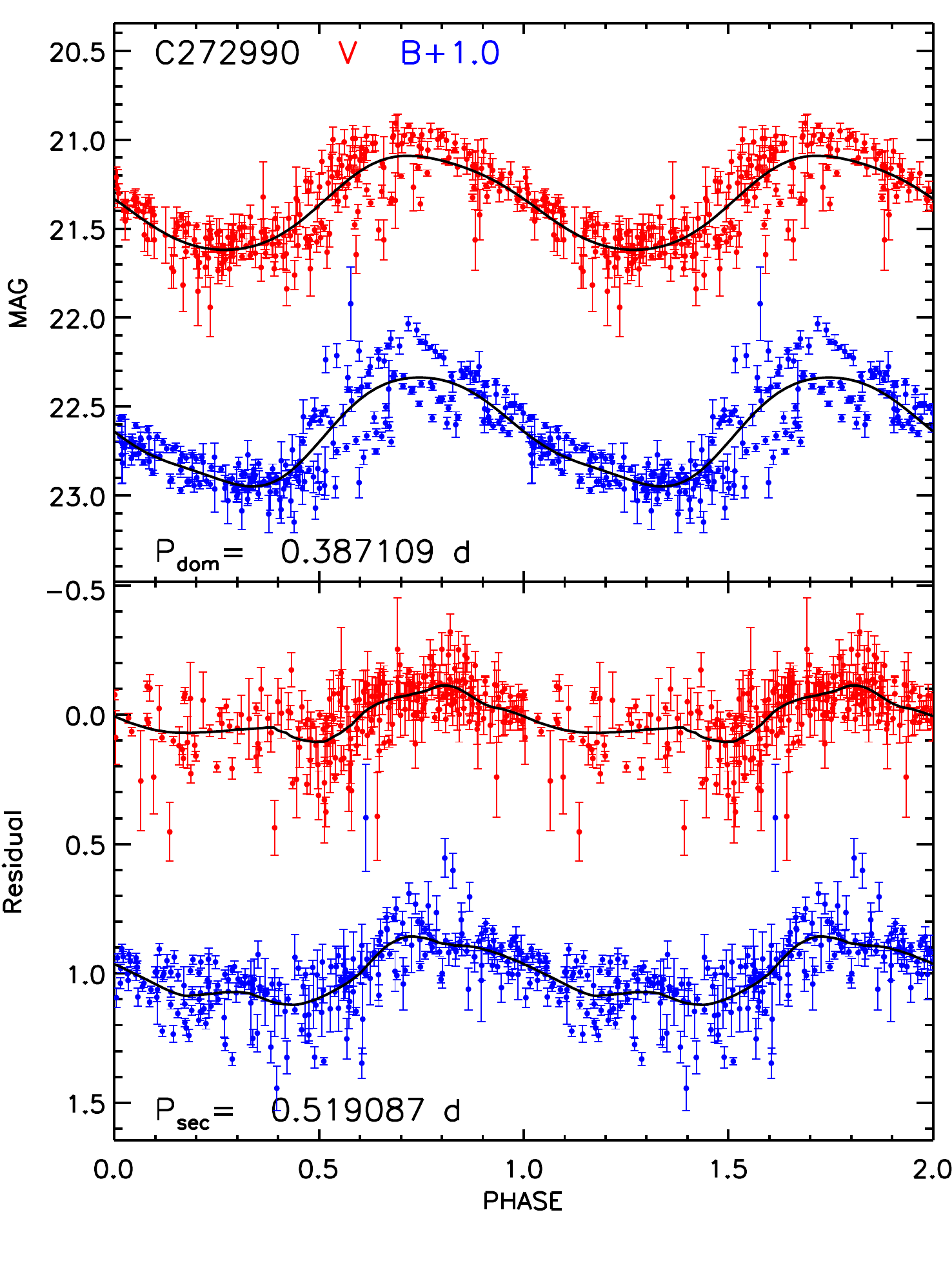}
    \caption{
    Top:  $V$- (red) and $B$-band (blue) light curves of the RRd C272990. The $B$-band light curve is shifted by 1 mag to improve visibility. The light curves are folded at $P_{dom}$, labelled at the bottom. The light curve fits are displayed as black solid lines.
    bottom: $V$- and $B$-band light curve of the residuals obtained by subtracting to the original data the fitting model of the dominant model from the empirical data. The light curve is folded at $P_{sec}$, labelled at the bottom. The light curve fits are displayed as black solid lines.
    }
    \label{fig:rrd_lcv}
\end{figure}

After this first separation, we fitted the folded $B$ and $V$ light curves of the remaining multiperiodic variable candidates (212 candidate RRds) with Fourier series and pre-whitened the time series by subtracting the model fit. Finally, we ran our periodicity search algorithm \citep[][]{braga2019b} on the residuals to estimate the period of the secondary mode ($P_{sec}$). Note that the adopted algorithms could not detect a clear periodic behavior for seven out of the 212 candidate mixed-mode RRLs. Based on their $P_{dom}$ and on a visual comparison with the light curves of literature bona-fide RRds, we kept these objects as candidate RRds, but we could not characterize their secondary mode. For the other 205 objects, we derived their period ratios ($P_{dom}$/$P_{sec}$) and---when possible---estimated the $V$- and $B$-band amplitudes associated to both primary and secondary periodicity ($Amp(B)_{dom/sec}$ and $Amp(V)_{dom/sec}$). Figure~\ref{fig:rrd_lcv} displays the $B$- and $V$-band light curves of the RRd C272990 detected in Fornax.


Finally, we classified the entire data set of RRLs into the subclasses RRab, RRc, RRd and Blazhko. The latter two were separated by following the procedure discussed at the beginning of this subsection. RRab and RRc were separated by means of their position on the Bailey diagram \citep[][and references therein]{clementini2022}. RRLs close to the separating line were visually inspected and manually classified as RRab or RRc according to the shape of their light curves (saw-tooth versus sinusoidal). We ended up with 2,068 RRLs, including 1,493 RRab, 363 RRc and 212 candidate RRd variables. We have also marked 181 RRLs of the 2068 RRLs as Blazhko candidates, and among them 179 are RRab and 2 are RRc variables. Recent estimates, based on OGLE-IV data, indicate that the fraction of RRab and RRc variables showing Blazhko modulations is 40.3\% \citep[][]{prudil2017} and 7.6\% \citep[][]{netzel2018}. The current fractions are significantly smaller, but this difference is expected. The number of measurements per variable in the OGLE-IV dataset is one-two orders of magnitude larger, thus suggesting that we are probably missing RRLs with small amplitude Blazhko modulations. Table~\ref{tbl:rrl_phot} lists the positions and the mean magnitudes of the 212 candidate RRds that we detected in Fornax.

\begin{table*}
\caption{Positions and mean magnitudes of the Fornax RRds. Only ten rows are listed. The table will be fully available in electronic format.}
\label{tbl:rrl_phot}
\begin{tabular}{l cc c c c c}
\hline
Name & RA & DEC & $<V>$ & $<B>$ & $<R>$ & $<I>$ \\
     & \multicolumn{2}{c}{deg} & \multicolumn{4}{c}{mag} \\
\hline
   A115625 &  39.564875 & --34.840639 &  21.323 $\pm$   0.026  &  21.611 $\pm$   0.024  &  21.143 $\pm$   0.059  &  20.925 $\pm$   0.031 \\
   A117688 &  39.572042 & --34.520583 &  21.408 $\pm$   0.028  &  21.646 $\pm$   0.025  &  21.089 $\pm$   0.048  &        \ldots         \\
   A119208 &  39.574000 & --34.829028 &  21.341 $\pm$   0.022  &  21.616 $\pm$   0.023  &  21.215 $\pm$   0.040  &  20.948 $\pm$   0.047 \\
   A235107 &  39.768792 & --34.116944 &  21.496 $\pm$   0.028  &  21.819 $\pm$   0.026  &  21.360 $\pm$   0.216  &  20.937 $\pm$   0.061 \\
   A241835 &  39.777833 & --34.138028 &  21.328 $\pm$   0.026  &  21.618 $\pm$   0.025  &  21.308 $\pm$   0.046  &  20.969 $\pm$   0.082 \\
   A286691 &  39.836750 & --34.072000 &  21.376 $\pm$   0.033  &  21.689 $\pm$   0.024  &  21.244 $\pm$   0.147  &  20.934 $\pm$   0.081 \\
   A311494 &  39.872208 & --34.020083 &  21.353 $\pm$   0.028  &  21.627 $\pm$   0.026  &  21.163 $\pm$   0.086  &  20.901 $\pm$   0.066 \\
   A322567 &  39.887500 & --33.960306 &  21.389 $\pm$   0.027  &  21.754 $\pm$   0.024  &        \ldots          &  21.003 $\pm$   0.070 \\
   A331006 &  39.899333 & --34.117556 &  21.442 $\pm$   0.029  &  21.743 $\pm$   0.026  &  21.290 $\pm$   0.041  &  21.099 $\pm$   0.103 \\
   A427173 &  40.018208 & --34.888556 &  21.336 $\pm$   0.026  &  21.518 $\pm$   0.019  &  21.074 $\pm$   0.101  &  21.054 $\pm$   0.098 \\
\hline
\end{tabular}
\end{table*}

\subsection{RRd identification in Sculptor}\label{sec:sculptor}

In order to compare RRds in Fornax and in Sculptor dSphs, we decided to revise recent RRd catalogs available in the literature for the latter system. To address this issue we adopted the most recent and extensive dataset of RRd light curves in Sculptor \citep[][]{martinezvazquez16b}.

We inspected the $B$- and $V$-band light curves of the 50 RRd candidates in \citep[][]{martinezvazquez16b} and the 18 (15 bona-fide plus three candidate) RRds in \citep[][]{kovacs2001}, moreover, we re-derived the periods by using the  
same approach adopted for Fornax (see Section~\ref{sect_ident}). In passing we also note that the current estimates of the $P_{dom}$ and $P_{sec}$ match those by \citep[][]{kovacs2001} up to the fourth decimal figure. The new estimates of $P_{dom}$ compared to \citet[][]{martinezvazquez15,martinezvazquez16b} match on average up to the sixth decimal figure, with the exception of three of them for which the difference is of the order of $10^{-2}$--$10^{-3}$ days. Note that we rejected 13 candidates from the sample of \citet[][]{martinezvazquez15,martinezvazquez16b} because the light curves of their residuals do not show solid secondary periodicities. These RRd candidates need to be investigated in more detail. We matched the two catalogs and we found that 16 objects overlap and 22 only in \citep[][]{martinezvazquez16b}, mostly thanks to the larger observed sky area. We ended up with a final catalog of 38 candidate RRds. Table~\ref{tbl:rrl_puls} lists the periods and light amplitudes of both Fornax and Sculptor RRds. Table~\ref{tbl:rrl_phot} does not list the Sculptor RRds because we did not redetermine the positions or magnitudes of the Sculptor RRds.  We refer the reader to \citet[][]{martinezvazquez15} for this information. 

\begin{table*}
\caption{Pulsation properties of Fornax and Sculptor RRds. The first column indicates whether the RRd belonging either to Fornax (F) or to Sculptor (S). The fourth column gives the dominant period: 0=FU; 1=FO. Only ten rows are listed. The table will be fully available in electronic format.}
\label{tbl:rrl_puls}
\begin{tabular}{ll cc r c c c c c c}
\hline
dSph & Name & $P_{dom}$ & $P_{sec}$ & Flag$_P$ & $Amp(V)$ & $Amp(B)$ & $Amp(R)$ & $Amp(I)$ & $Amp(V)_{sec}$ & $Amp(B)_{sec}$ \\
    & & \multicolumn{2}{c}{days} & & \multicolumn{6}{c}{mag} \\
\hline
   F & A115625 &   0.36726727 &   0.45028790 & 1 &   0.61 $\pm$   0.07 &   0.76 $\pm$   0.06 &       \ldots       &       \ldots       &       \ldots       &   0.19 $\pm$   0.03 \\
   F & A117688 &   0.37921800 &   0.89228848 & 1 &   0.56 $\pm$   0.05 &   0.68 $\pm$   0.06 &   0.64 $\pm$   0.34 &       \ldots       &       \ldots       &   0.19 $\pm$   0.02 \\
   F & A119208 &   0.44559100 &   0.61226688 & 1 &   0.38 $\pm$   0.04 &   0.45 $\pm$   0.04 &       \ldots       &       \ldots       &   0.20 $\pm$   0.04 &   0.21 $\pm$   0.03 \\
   F & A235107 &   0.39703400 &   0.38717016 & 1 &   0.32 $\pm$   0.04 &   0.38 $\pm$   0.04 &       \ldots       &       \ldots       &       \ldots       &       \ldots       \\
   F & A241835 &   0.39947200 &   0.53613751 & 1 &   0.50 $\pm$   0.06 &   0.68 $\pm$   0.06 &       \ldots       &       \ldots       &   0.30 $\pm$   0.05 &   0.42 $\pm$   0.04 \\
   F & A286691 &   0.41394600 &   0.55573180 & 1 &   0.26 $\pm$   0.04 &   0.41 $\pm$   0.04 &       \ldots       &       \ldots       &   0.28 $\pm$   0.06 &   0.30 $\pm$   0.03 \\
   F & A311494 &   0.37354600 &   0.51045320 & 1 &   0.36 $\pm$   0.05 &   0.66 $\pm$   0.05 &       \ldots       &       \ldots       &   0.24 $\pm$   0.04 &   0.37 $\pm$   0.03 \\
   F & A322567 &   0.35348900 &   0.46174616 & 1 &   0.66 $\pm$   0.07 &   0.74 $\pm$   0.04 &       \ldots       &       \ldots       &       \ldots       &   0.25 $\pm$   0.03 \\
   F & A331006 &   0.36132912 &   0.47121154 & 1 &   0.39 $\pm$   0.05 &   0.56 $\pm$   0.05 &       \ldots       &       \ldots       &   0.48 $\pm$ -99.00 &       \ldots       \\
   F & A427173 &   0.40955600 &   0.54764900 & 1 &   0.26 $\pm$   0.04 &   0.56 $\pm$   0.05 &       \ldots       &       \ldots       &       \ldots       &   0.18 $\pm$   0.03 \\
\hline
\end{tabular}
\end{table*}

\section{Old and new pulsation diagnostics for RRds}\label{sec:diagnostics}

To constrain the properties of RRds, we will adopt the following diagnostics: the fraction of RRds with respect to the total number of RRLs ($\eta^{RRd}_{RRL}=\dfrac{N_{RRd}}{N_{RRL}}$), the PD ($P_1$/$P_0$ vs $P_0$) and a new diagram that we named Period-Amplitude RatioS (PARS).

To provide a comprehensive scenario of the pulsation properties of RRds we extend our analysis to four GGCs, (IC4499, NGC4590, NGC5272, NGC7078) three dSphs (Sagittarius, Draco, Carina), the MCs and the Bulge. Note that we adopted the Bulge sample from OGLE-IV \citep[][]{soszynski2019} and this sample includes both Bulge RRLs and RRLs belonging to the inner Halo. For simplicity, in the rest of the paper, we will call this simply the ``inner galaxy'' (IG) sample.
Table~\ref{tab:dsph} lists the properties of these stellar systems. Note that only 32 out of the 38 candidate Sculptor RRds and 130 out of 212 candidate Fornax RRds are located within the main sequence of the PD (see Section~\ref{sec:petersen}). We found that the outliers suffer for poor/limited light curve coverage, while those in the main-sequence, that we consider as ``canonical RRds'' are better sampled. We will only consider this subsample of canonical RRds in the rest of the paper. For the same reason we will not include stars flagged as ``anomalous RRd'' in the OGLE data sets \citep[][]{soszynski2019,soszynski2019c}.
We point out that, despite the catalogs of RRd were collected from heterogeneous samples and from systems with different metallicities, there is no empirical/theoretical evidence that the chemical composition can affect the completeness of the sample due to misclassification. This might be an issue when dealing with RRd catalogs based on investigation with no specific analysis and focus on RRds identification \citep[][]{kaluzny95} where, according to \citep[][]{kovacs2001}, a large fraction of RRds were classified as RRc. Note that the possible incompleteness in the number of RRds goes in the direction to possibly increase and not to decrease the current estimates of the population ratio.

\begin{table*}
\caption{Properties of the stellar systems hosting RRds. From left to right: the stellar system, the number of RRab, and RRd stars, the iron abundance and its standard deviation. For Sculptor and Fornax, we indicate both the number of candidate and bona-fide RRds. The seventh column gives the references for $N_{RRL}$ and $N_{RRd}$: 
1=\citep[][]{soszynski2019c}; 
2=\citep[][]{soszynski2019};
3,4=\citep[][]{hamanowicz2016,figuerajaimes2016};
5=This work;
6=\citep[][]{coppola15};
7=\citep[][this work]{martinezvazquez15};
8,9=\citep[][]{kinemuchi08,muraveva2020}
10,11,12=\citep[][]{corwin2008,hoffman2021,bhardwaj2021};
13=\citep[][]{kains2015};
14=\citep[][]{walker1996};
15=\citep[][and references therein]{jurcsik2019}.
The eighth column lists the references for the iron abundance: 
1=\citep[][]{gratton04a}; 
2=\citep[][]{skowron2016};
3=\citep[][]{walker1991a};
4=\citep[][]{alfarocuello2019};
5=\citep[][]{lemasle14};
6,7=\citep[][]{monelli14,fabrizio15};
8=\citep[][]{martinezvazquez16a};
9=\citep[][]{kirby2015}
10=\citep[][]{carretta09}.}
\label{tab:dsph}
\begin{tabular}{l rrrcccc}
\hline
Name & $N_{RRab}$ & $N_{RRc}$ & $N_{RRd}$ & $\eta_{RRd}$ & [Fe/H] & Ref.  & Ref. \\
     &           &           &           &    &    & $N$ & [Fe/H] \\
\hline
Sagittarius  &  140 &  32   &    3     & 0.017& --1.41$\pm$0.24 & 3,4 & 4 \\
Sculptor     &  293 & 205   &   32/38  & 0.060& --1.84$\pm$0.34 & 7 & 8 \\
Draco        &  224 &  35   &   26     & 0.091& --2.05$\pm$0.37 & 8,9 & 9 \\
Carina       &   71 &  12   &   9      & 0.098& --2.13$\pm$0.28 & 6 & 6,7 \\
Fornax       & 1493 & 363   &  130/212 & 0.065& --2.10$\pm$0.27 & 5 & 5 \\
IG           &48662 & 19072 & 333      & 0.005& --1.05$\pm$0.16 & 2 & 3 \\
LMC          &29028 & 834   & 2083     & 0.051& --1.48$\pm$0.29 & 1 & 1 \\
SMC          & 5201 & 9985  & 674      & 0.100& --1.77$\pm$0.48 & 1 & 2 \\
NGC 7078     &   67 &  69   &   29     & 0.176& --2.33$\pm$0.02 & 10,11,12 & 10 \\
NGC 4590     &   14 &  16   &   12     & 0.286& --2.27$\pm$0.04 & 13 & 10 \\
IC 4499      &   67 &  13   &   17     & 0.175& --1.62$\pm$0.09 & 14 & 10 \\
NGC 5272     &  241 &  48   &   10     & 0.041& --1.50$\pm$0.05 & 15 & 10 \\
\hline
\end{tabular}
\end{table*}



\subsection{Number fraction of RRds}\label{sec:fraction}

On the basis of OGLE-III data of RRds in the MCs and in the IG, \citep[][]{soszynski2011,soszynski14} plus similar data for GGCs and nearby dSphs, \citet{coppola15} suggested that the ratio of RRd over the total number of RRLs is anti-correlated with metal abundance. This finding is also consistent with the theoretical predictions because an increase in metal content causes the HB morphology to become systematically redder. This means that the hotter regions of the instability strip, in which are present RRc and RRd variables, are either minimally or not populated \citep{bono97b,szabo04}. OGLE-IV data, together with the new and catalogs for RRds of dSphs and GGCs allow us to investigate this difference by using significantly larger datasets.


To investigate the properties of RRds on a variety of stellar systems covering a broad range in metallicities, we estimated $\eta_{RRd}$ in the systems listed in Table~\ref{tab:dsph}. Note that LMC, SMC, Bulge and Sagittarius samples are all based on OGLE-IV data, and data reduction and analysis are the same. Concerning Sagittarius RRds, we also employed a complementary catalog \citep{figuerajaimes2016} based on EMCCD@Danish data that, for the variables in common, agrees quite well with those based on OGLE-IV \citep{hamanowicz2016}. The remaining dSph and GGC catalogs come from investigations focussed specifically on RRds and based on photometric time series including from  tens to hundreds of phase points.


On the basis of the new OGLE-IV data \citep{soszynski2019c,soszynski2019}, we estimated $\eta_{RRd}$ in the IG and in the MCs and we found, in agreement with \citep[][]{coppola15}, that $\eta_{RRd}$ increases by a factor of twenty when moving from metal-rich ($\eta_{RRd}$(IG)=0.49\%) to more metal-poor ($\eta_{RRd}$(LMC)=5.1\%, $\eta_{RRd}$(SMC)=10.0\%) stellar systems. We found that $\eta^{RRd}_{RRab}$ follows the same behavior. By counting also the anomalous RRds---those for which the dominant mode is the FU---we found that $\eta_{RRd}$(IG) increases to 0.56\% (a $\sim$10\% increase), while for the LMC and SMC, adding the anomalous RRds, leaves $\eta_{RRd}$ unchanged. This means that metallicity might have a small effect on the ratio of anomalous RRds. Note that a similar comparison with RRds in other stellar systems is hampered either by statistics or by the lack of homogeneous and accurate data sets. The main aim of this analysis is to investigate whether the population ratio changes when moving from metal-poor to metal-rich stellar systems. A detailed quantitative analysis of this trend is out of the aim of the current investigation.

\begin{figure}
	\includegraphics[width=\columnwidth]{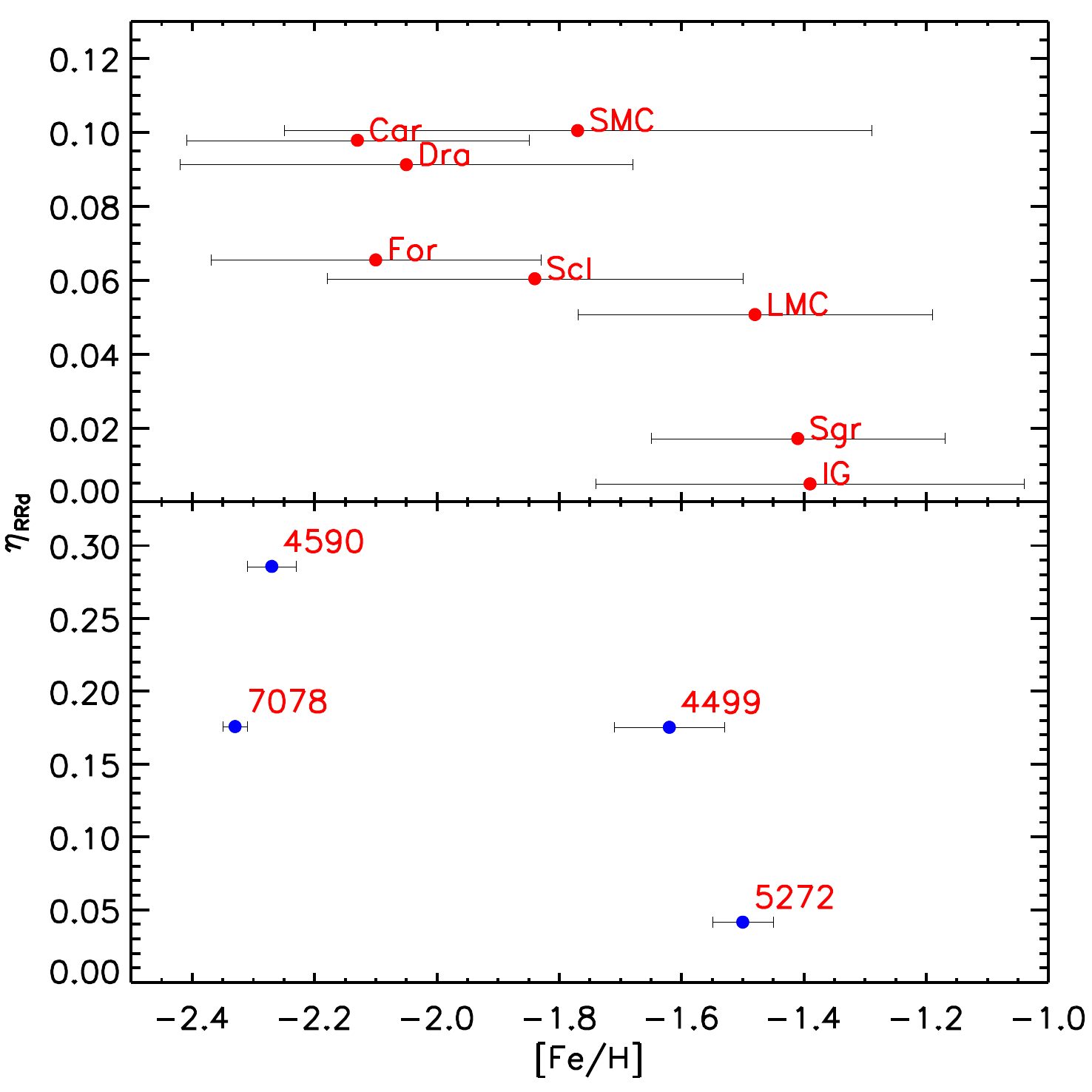}
    \caption{Top: Number fraction of RRds ($\eta^{RRd}_{RRL}=\dfrac{N_{RRd}}{N_{RRL}}$) versus iron abundance of the old stellar population for different stellar systems. The bars mark the standard deviations of the [Fe/H] distributions of old stars in these systems. 
    Bottom: The GGCs are marked as blue circles.}
    \label{fig:eta_vs_feh}
\end{figure}


We have also derived $\eta_{RRd}$ for several dSph and GGCs and their values are listed in Table~\ref{tab:dsph}. A recent investigation \citet[][]{cseresnjes2001} provided a more complete sample of Sagittarius RRds (40 RRds) by using  a set of photographic plates covering an area of 50 square degrees, but this investigations lacks the characterization of the RRL sample, thus hampering the estimate of the population ratio. Moreover, the estimate of $\eta_{RRd}$ for Fornax and Sculptor dSphs is a lower limit because---as mentioned in Section~\ref{sec:fornax} and~\ref{sec:sculptor}---there are several more RRd candidates which could not be classified as bona-fide RRds. Note that Halo RRds were not included in the current analysis because their classification is based on the the \texttt{best\_classification} column provided by Gaia DR3 \texttt{gaiadr3.vari\_rrlyrae}, which is still prone to possible systematics for RRds \citep[][see fig. 5]{clementini2022}.

Since the stellar systems that we discuss show quite complex star formation histories, we need to pay some attention to constrain the stellar metallicity of their old stellar population. When available, we have selected [Fe/H] estimates based on RRL themselves. This is the case of LMC (high-resolution spectroscopy), SMC (photometric metallicities) and IG ($\Delta$S method from low-resolution spectra), see the references listed in Table~\ref{tab:dsph}. We have also favored spectroscopic over photometric metallicity estimates. By following these criteria, we selected the [Fe/H] values provided in Table~\ref{tab:dsph}.
The inclusion of the dSphs to Figure~\ref{fig:eta_vs_feh} shows that the mean difference between metal-poor and metal-rich stellar systems is confirmed in spite 
of the large dispersion of both $\eta_{RRd}$ at [Fe/H] lower than $\sim$--2.0 dex. The quoted dispersion is mostly due to the uncertainty on the precise number of RRds in Fornax and in Sculptor. 

This trend is supported by theoretical predictions: \citet{bono97b,szabo04} found that the so-called "OR Region", i.e. the region of the Instability Strip in which both fundamental and first overtone are simultaneously stable, is only marginally crossed by the more metal-rich HB evolutionary models. Moreover and even more importantly, the "OR region" becomes systematically narrower when moving from the metal-poor to the metal-rich regime \citep[][]{marconi15}
The bottom panel of Figure~\ref{fig:eta_vs_feh} shows $\eta_{RRd}$ versus [Fe/H] for four GGCs. The analysis is hampered by the limited statistics. However, the metal-poor Oosterhoff II cluster NGC~4590 ([Fe/H]$<$--2, $\dfrac{N_{RRc}}{N_{RRL}}>$ 0.55) has a population ratio that is a factor of seven larger than for the metal-intermediate Oostehoff I cluster NGC~5272 (0.286 vs 0.041). On the other hand, the two clusters NGC~7078 and IC~4499 have a difference in metallicity of $\sim$0.7 dex, but they have similar population ratios. Finally,  IC~4499 and NGC~5272 are Oosterhoff I clusters, but they show a difference of a factor of four in $\eta_{RRd}$.
Note that the comparison between RRds in GGCs and in dSphs is far from being obvious. The HB morphology and the sampling of the RRL instability strip in GCs are affected by the metallicity and by the so-called second parameter problem \citep[][and references therein]{torelli2019}. Moreover, they are typically affected by small number statistics. On the other hand, the horizontal portion of the HB, and in turn of the IS, is well populated in nearby dSphs and RRLs have been identified in all the investigated systems.

\subsection{Petersen Diagram}\label{sec:petersen}
\begin{figure}
	\includegraphics[width=\columnwidth]{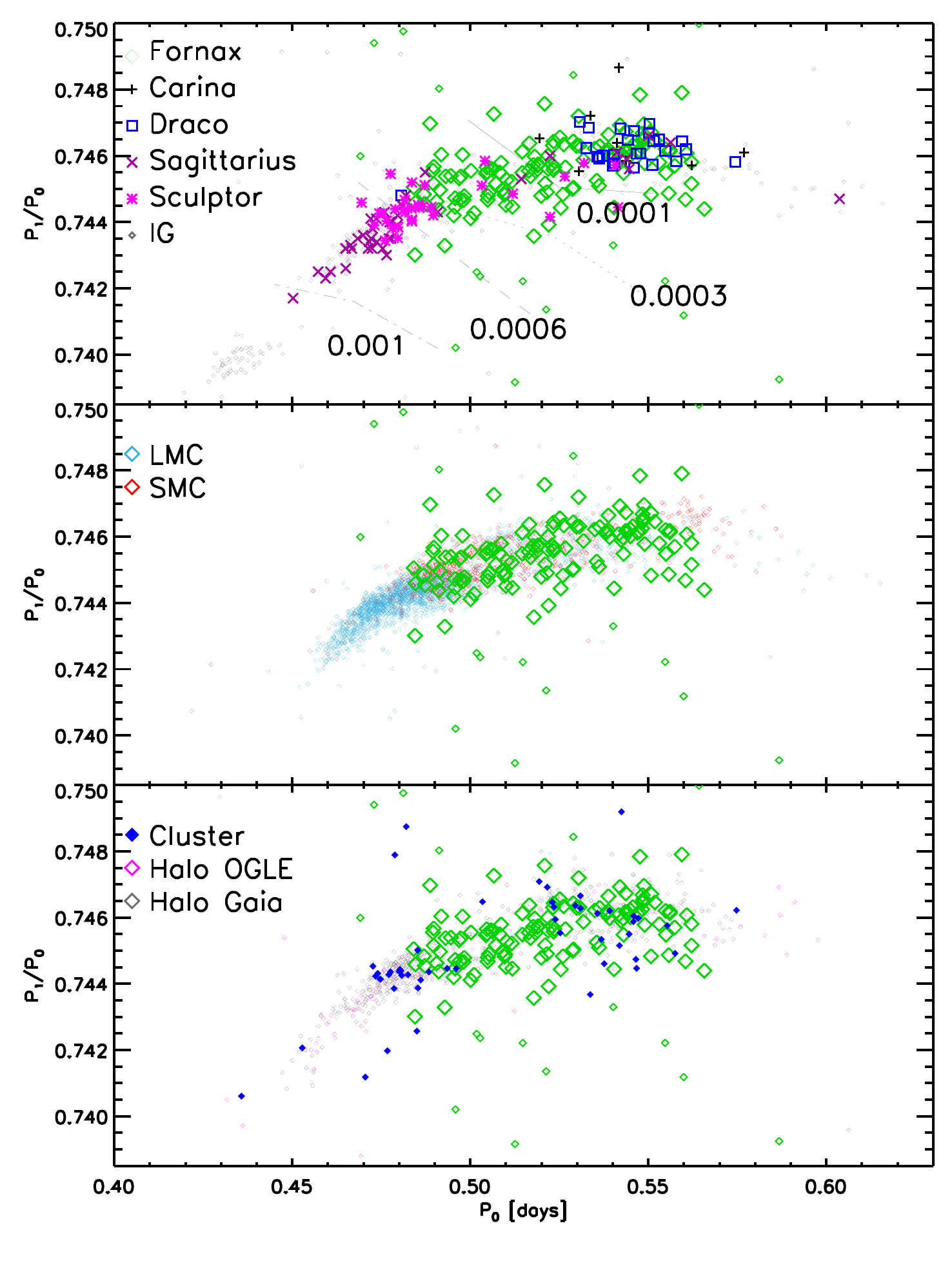}
    \caption{Petersen diagram: period ratio ($P_1$/$P_0$) versus the fundamental period ($P_0$) for the current RRd sample. 
    Top: RRds in dSphs and in the IG. Green diamonds, Fornax; black pluses, Carina; blue squares, Draco; purple crosses, Sagittarius; magenta asterisks, Sculptor; small grey diamonds, IG. The pulsation models for RRds at different metal contents \citep[][]{marconi15} are displayed as solid, dotted, dashed and solid-dashed lines. The metal abundances (mass fraction) are labeled.
    Middle: Same as the top, but for RRds in the LMC (light-blue diamonds) in the SMC (red diamonds) and in Fornax. 
    Bottom: Same as the top, but for RRds in Fornax, in GGCs (blue diamonds, \citealt[][and references therein]{Clement01}) and in the Halo based on the OGLE (magenta) and on the \textit{Gaia} (black diamonds) catalogs \citealt{soszynski2019,clementini2019}).}
    \label{fig:petersen}
\end{figure}

The dominant pulsation mode of RRds is the FO and the majority of the RRds are located within a very narrow region in the instability strip \citep[][]{cox1983}. This is clearly visible from the limited range in pulsation periods covered by RRds in the LMC and in the SMC (see middle panel in Figure~\ref{fig:petersen}). As already pointed out by \citet{bono1996} and \citet{popielski2000} and addressed on an empirical basis by \citet{soszynski2011} and \citet{coppola15}, RRds are ranked by metallicity in the PD. More precisely, $P_1$/$P_0$ and $P_0$ steadily decrease when moving from more metal-poor to more metal-rich stellar systems. The decrease in $P_0$ and in $P_1$ as a function of the iron abundance has been soundly confirmed  by using the largest spectroscopic sample of field RRLs (6150 RRab, 2865 RRc) ever collected \citep[][]{fabrizio2019,fabrizio2021}. The quoted trends take account of the metallicity trends observed in the PD. 

This effect is clearly visible in the top panel of Figure~\ref{fig:petersen} showing the distribution of RRds in dSphs. The ranking in metallicity is supported both by the empirical metallicity estimates listed in Table~\ref{tab:dsph} and by the comparison with pulsation models \citep[][]{marconi15}. The current empirical evidence indicates that an increase in metal content causes a steady decrease in the period ratio and in the fundamental period.  
Note that, in Figure~\ref{fig:petersen}, only the ZAHB models are displayed. The general trend of the models with metallicity agrees with observations. Note that an increase in the metal content by one order of magnitude, from Z=0.0001 to 0.001 causes a decrease of $\sim$0.0035 in the period ratio and a decrease in $P_0$ of $\sim$0.07 days. The empirical data show the same trend and very similar numbers: according to the [Fe/H] values listed in Table~\ref{tab:dsph}, the difference in metallicity between Draco and Sagittarius\footnote{Note that for the Sagittarius RRds, we used data from \citet[][]{cseresnjes2001} because they provided pulsation properties for a sample of 40 RRds.} is also around one order of magnitude and the decrease in period ratio and in $P_0$ is $\sim$0.0035 and $\sim$0.08 days, respectively.

%
\begin{figure}
	\includegraphics[width=9.5cm]{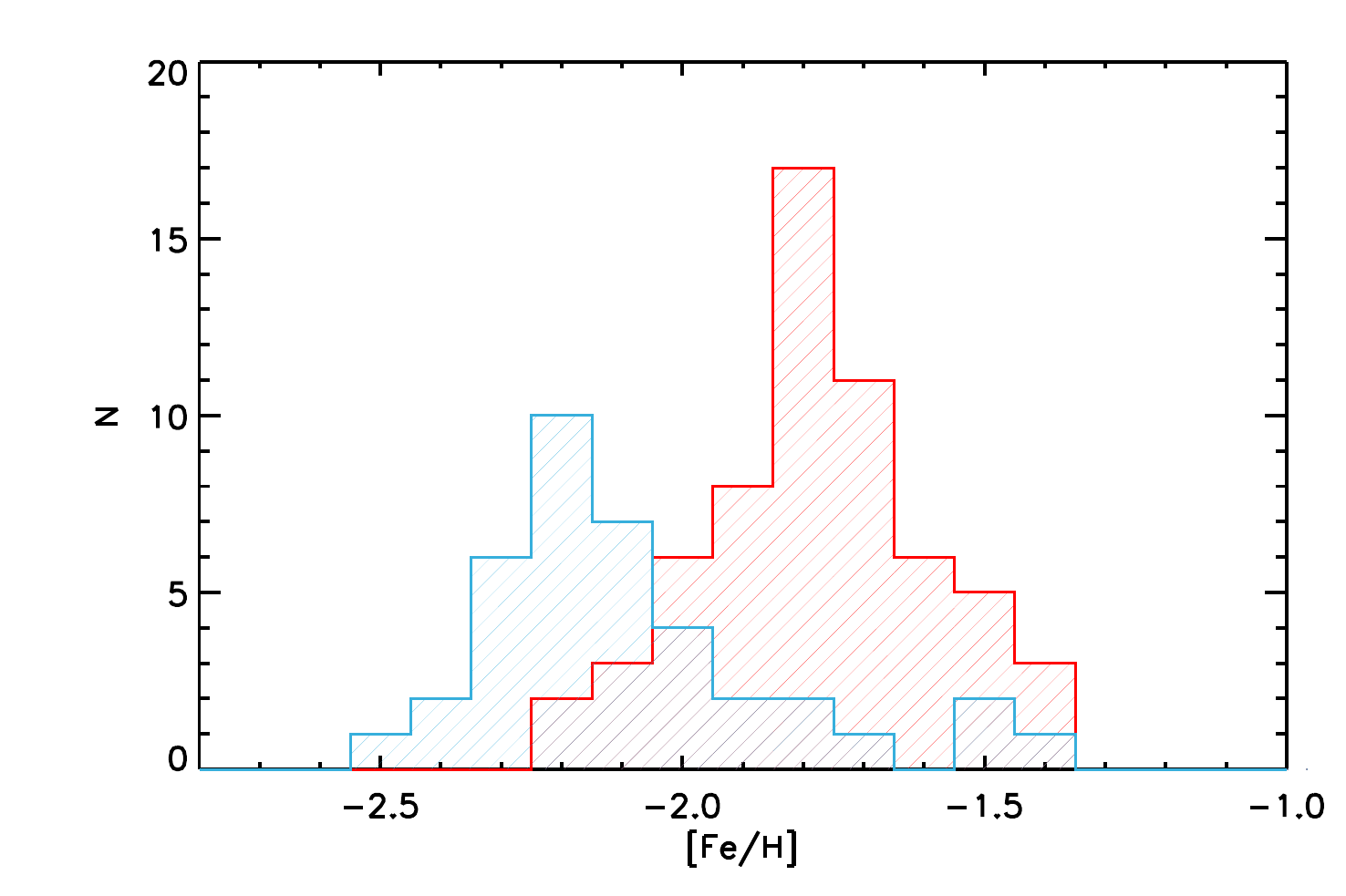}
    \caption{Metallicity distribution for Halo RRds derived by using iron abundances from \citet[][]{crestani2021a}. RRd stars are identified using Gaia DR3 \citep[][Fig]{clementini2022}. The blue and red bars show the distribution of the long- and short-period RRds, respectively.}
    \label{fig:rrd_histofe}
\end{figure}

To further constrain on a quantitative basis the ranking in metallicity across the PD, we investigated the iron distribution of Halo RRds. We matched the \textit{Gaia} catalog of Halo RRds with our own spectroscopic catalog of Halo RRLs \citep[][]{crestani2021a,fabrizio2021}  and we found 83 RRds (as classified in the \texttt{best\_classification} column of the Gaia DR3 catalog of RRLs) with iron abundances based on the new calibration of the $\Delta$S method. The median [Fe/H] of short-period RRds (see Section~\ref{sec:specialdiagram} for a definition of short- and long-period RRds) is --1.71$\pm$0.16 dex while for the long-period RRds is --2.07$\pm$0.22 dex (see Fig.~\ref{fig:rrd_histofe}).

\begin{figure*}
	\includegraphics[width=18.5cm, height=12truecm]{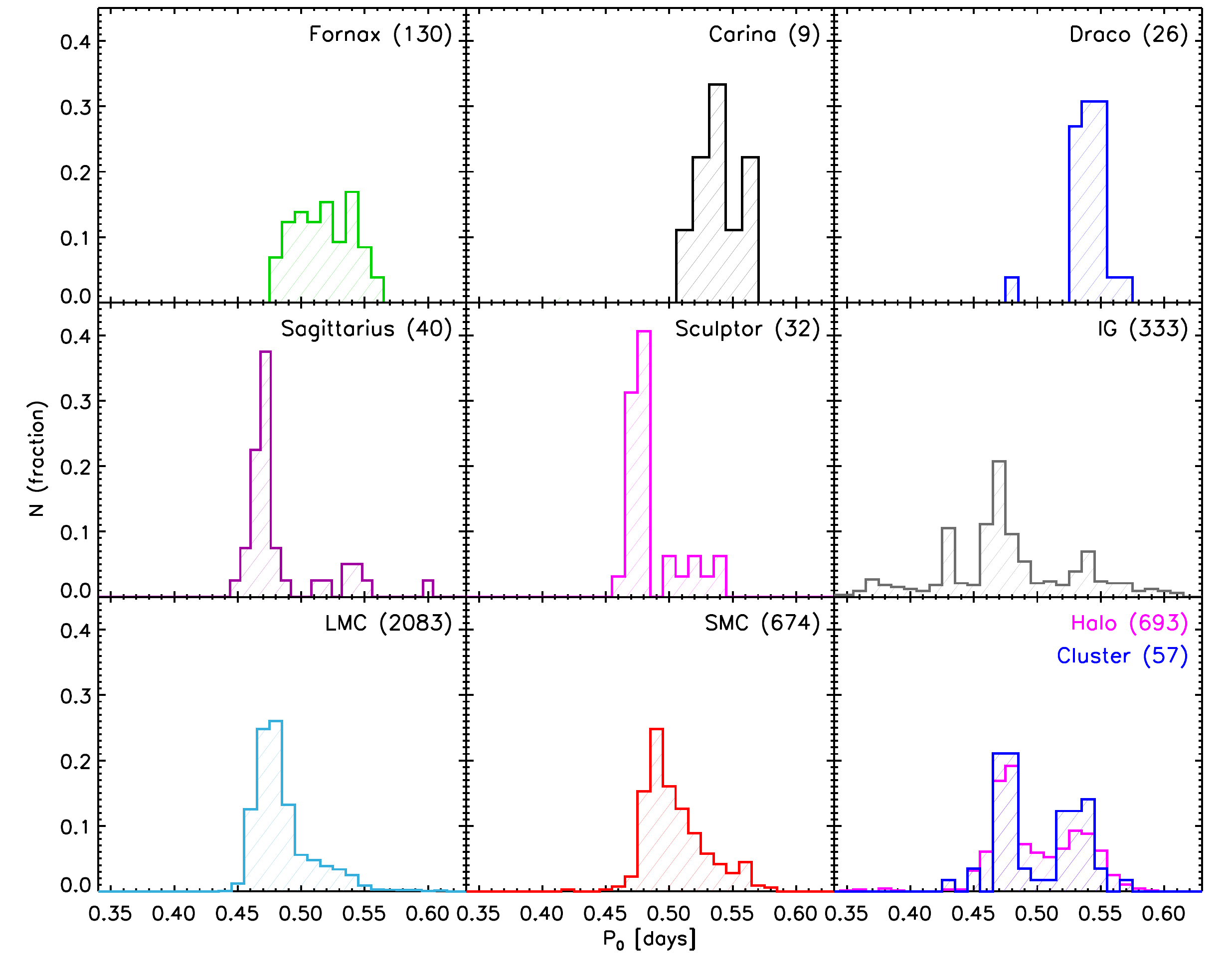}
    \caption{$P_0$ distribution of RRds in dSphs, MCs, the IG, Halo and GCs. Each panel corresponds to a galaxy (or galactic environment), with the only exception of Halo and GC RRds that are shown in the same panel. The number of RRds in each sample is labelled at the top-right corner of each panel.}
    \label{fig:histoper}
\end{figure*}

Note that RRds in Draco and Carina cover a small range in $P_0$, while RRds in Sagittarius, Sculptor and in Fornax cover a more extended region of the PD (see also the histograms in Fig.~\ref{fig:histoper}). More specifically, for Sculptor and Sagittarius, the majority of RRds clusters at short $P_0$, but a non-negligible fraction, presumably the more metal-poor component, extends up to $\sim$0.55 days. This evidence supports the broad metallicity distribution found by \citep[][]{martinezvazquez16a} by using the Period-Luminosity-Metallicity relation as a photometric index to estimate the metal content of individual RRLs. On the other hand, the RRds in Fornax are more uniformly distributed (top-left corner of Fig.~\ref{fig:histoper}), thus suggesting that the metallicity distribution of Fornax RRLs is wider and more uniform than in other nearby dSphs. This is supported by the age-metallicity relation found for Red Giant Branch stars (RGBs) in Fornax by \citet[][]{lemasle14}. They found that RGBs older than 10 Gyr display a very broad distribution in iron abundance (--2.5$\lesssim$[Fe/H]$\lesssim$--0.5). Furthermore, the new RRds that we found in Fornax and the updated catalog of RRds in Sculptor, fill a gap in fundamental period ($0.49<P_0<0.53$ days) that was not covered by RRds in dSphs available in the literature. 


The fact that dSph RRds do not display anymore a gap in period, is important when comparing dSph and Halo RRds. Indeed, Halo RRds cover, without gaps, the entire $P_0$ range from $\sim$0.44 to $\sim$0.59 days and the $P_0$ gap in dSph RRds was the missing interval when trying to link Halo and dSph RRds. We also note that both Fornax and Halo RRds display a flattening of $P_1/P_0$ at $P_0\sim$0.54 days. Unfortunately, the dSph sample is not complete, both due to the incompleteness of the individual catalogs and for the lack of data for several dSph (e.g. Ursa Minor, Sextans). The current data do not allow us to reach firm conclusions concerning the role that dSphs played in building up the Halo. 
However, the short-period, high-metallicity tail is only observed in massive dwarf galaxies, SMC, LMC and Sgr dSph, in analogy with what happens for the short period tail in the RRab distribution. The lack of High Amplitude Short Period (HASP) RRLs, was considered by \citet{fiorentino15a} a strong evidence of the limited role that less massive dwarf galaxies (including Fornax) played in the Halo assembly. They take account for, at most, the $\sim$20\% of the Halo mass \citep[see also][]{fiorentino17}.   

\begin{figure*}
	\includegraphics[width=8.5cm]{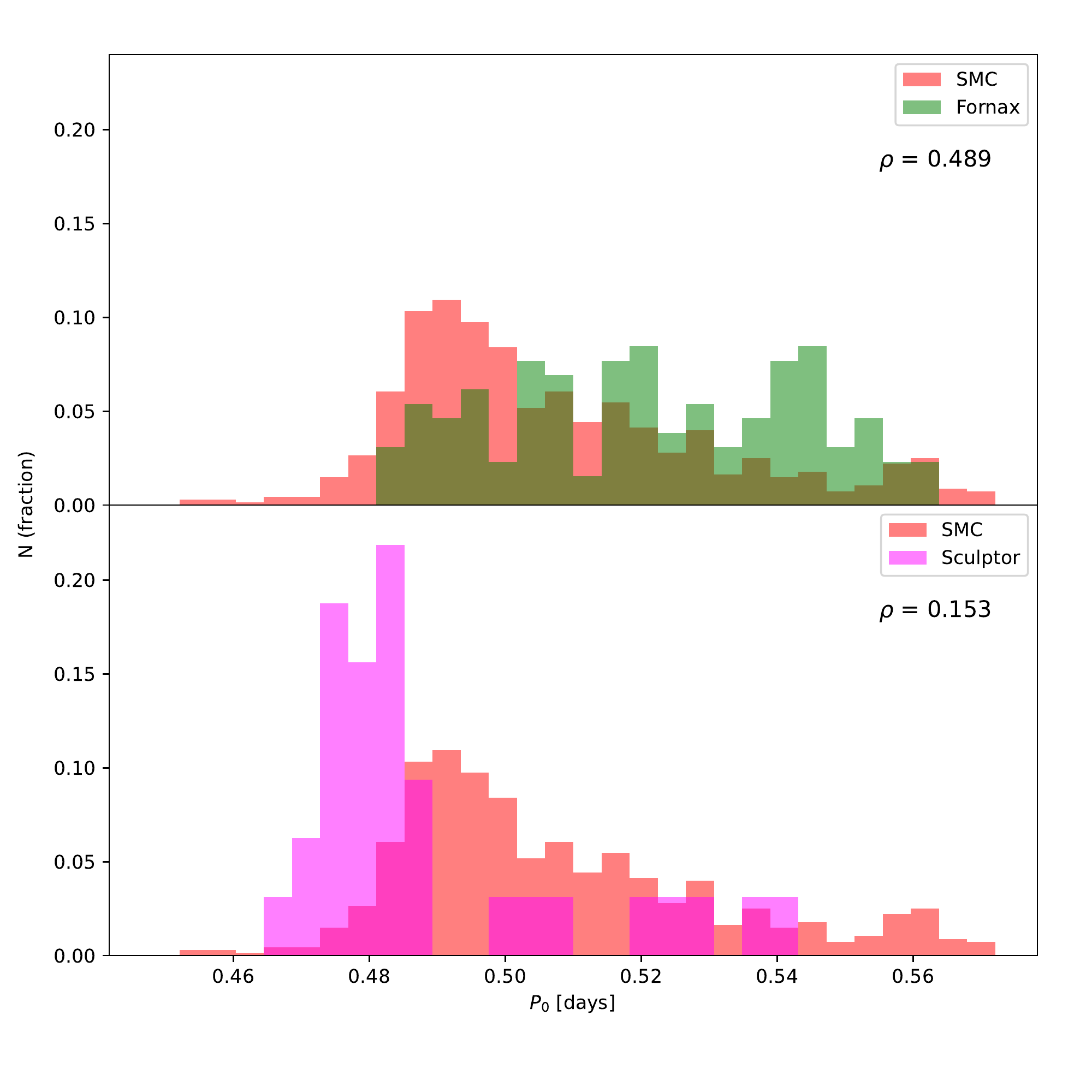}
	\includegraphics[width=8.5cm]{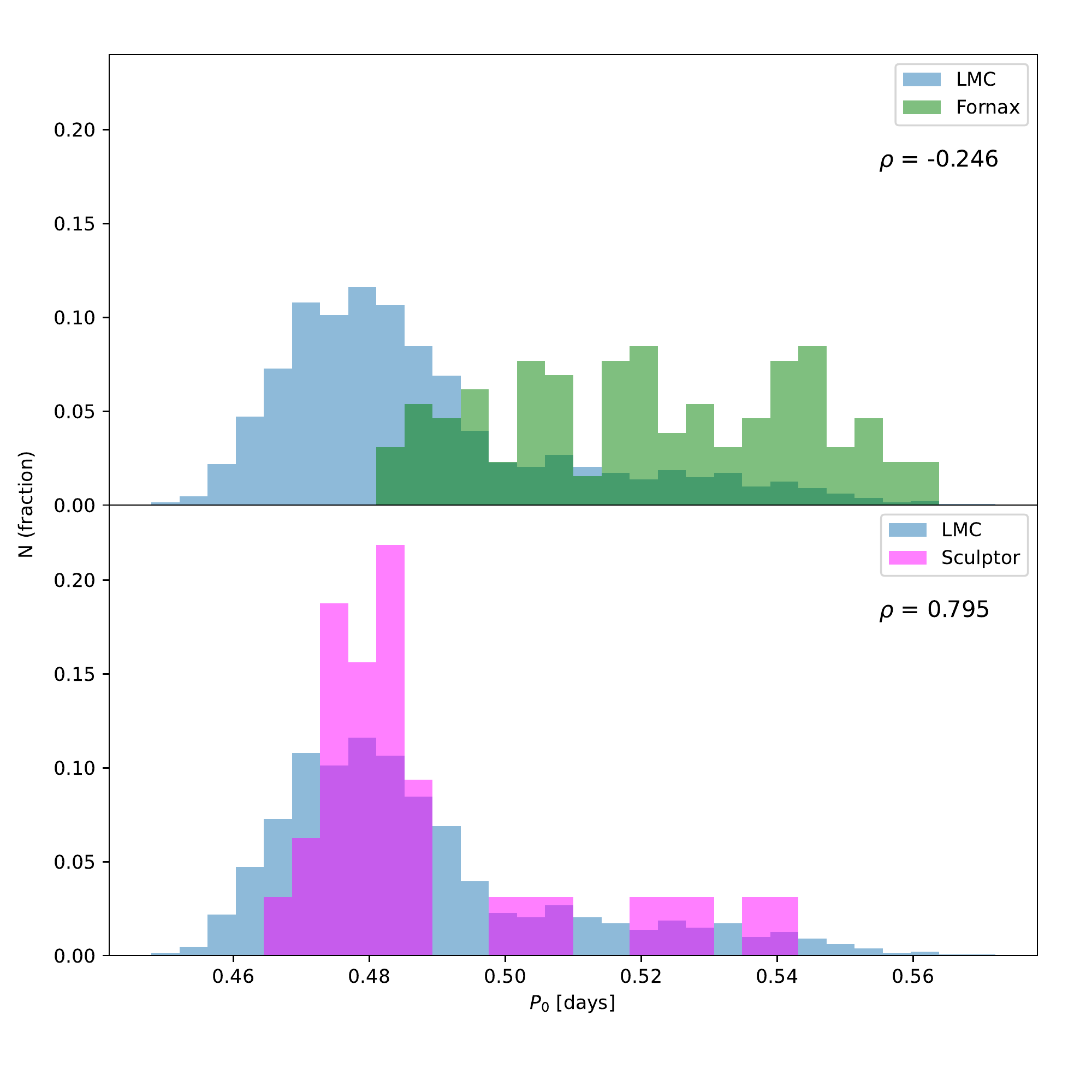}
    \caption{Top left: $P_0$ distribution of RRds in SMC (light blue) and Fornax (green); Bottom left: same as top but magenta for Sculptor; Top right: $P_0$ distribution of RRds in LMC (light blue) and Fornax (green); Bottom right: same as top but magenta for Sculptor.}
    \label{fig:giacinto}
\end{figure*}

To perform a quantitative analysis of the similarity among the period distributions of different stellar systems, we calculated the Pearson’s linear correlation coefficient ($\rho$) between the histograms of the LMC, SMC, Fornax and Sculptor. For all the possible pairs, we found the following values of the coefficient: LMC-Fornax= --0.246; LMC-Sculptor=0.795; SMC-Fornax=0.489; SMC-Sculptor=0.153. These estimates indicate that there is similarity only between LMC and Sculptor, while SMC and Fornax are only marginally similar (see Figure~\ref{fig:giacinto}). Furthermore, data plotted in the middle panel of Fig.~\ref{fig:petersen} suggest a similarity between RRds in Fornax and in the SMC. Indeed, the RRds in these two dwarf galaxies cover a similar period range of $0.48 \ltsim P_0 \ltsim 0.565$ days, suggesting that the metallicity distribution of their old stellar populations are also similar. In fact, according to the values in Table 3, they overlap over 0.5 dex (from --2.3 to --1.8) within less than 1$\sigma$. However, the distribution of the RRds in the SMC is more peaked and skewed towards shorter periods suggesting that, on average, its RRLs are slightly more metal-rich than in Fornax. This finding agrees quite well with the metallicity distributions recently provided by \citet[][]{skowron2016} by using metallicity estimates based on Fourier parameters of $I$-band light curves.


On the other hand, according to the Pearson’s coefficients, the similarity between Sculptor and the LMC is stronger. In this case, the Petersen diagram is not a solid diagnostic: indeed the two distributions appear different, since the short-period tail of the LMC that does not overlap with Sculptor. However, the short period tail is poorly populated (see, e.g., Fig.~\ref{fig:histoper}), thus suggesting that its weight, in the comparison of the two distributions is minimal. In Section~\ref{sec:specialdiagram}, we discuss in more detail the similarities among the quoted stellar systems.



The comparison with Halo RRds is made possible only thanks to the new identifications provided by the OGLE team and by Gaia. Indeed, the number of field RRds increases by a factor of two when compared with the number of RRds available in the GCVS. Also, GCVS does not provide the secondary period, thus hampering the analysis of the properties of RRds. The data plotted in the bottom panel of Fig.~\ref{fig:petersen} show that Halo RRds cover a broad range in pulsation periods ($0.45 \ltsim P_0 \ltsim 0.58$ days). However, the period distribution has two peaks: a primary one at $P_0$=0.48 days and a secondary one at 0.54 days, i.e. in the typical range of metal-intermediate/metal-poor RRLs. Note that the bi-modal distribution  is not connected with the Oosterhoff dichotomy. Indeed, dSphs are Oosterhoff intermediate and they fill the gap between OoI and OoII GGCs \citep{catelan09,braga16}. Moreover, the average periods of the two RRd groups, both $P_0$ and $P_1$, differ from the average periods of the Oosterhoff groups.

The period histograms plotted in Fig.~\ref{fig:histoper} suggest that both the Halo and GC RRds display a well defined dichothomy, while the IG seems to show a broad, multimodal distribution, with three peaks.

\begin{figure}
	\includegraphics[width=\columnwidth]{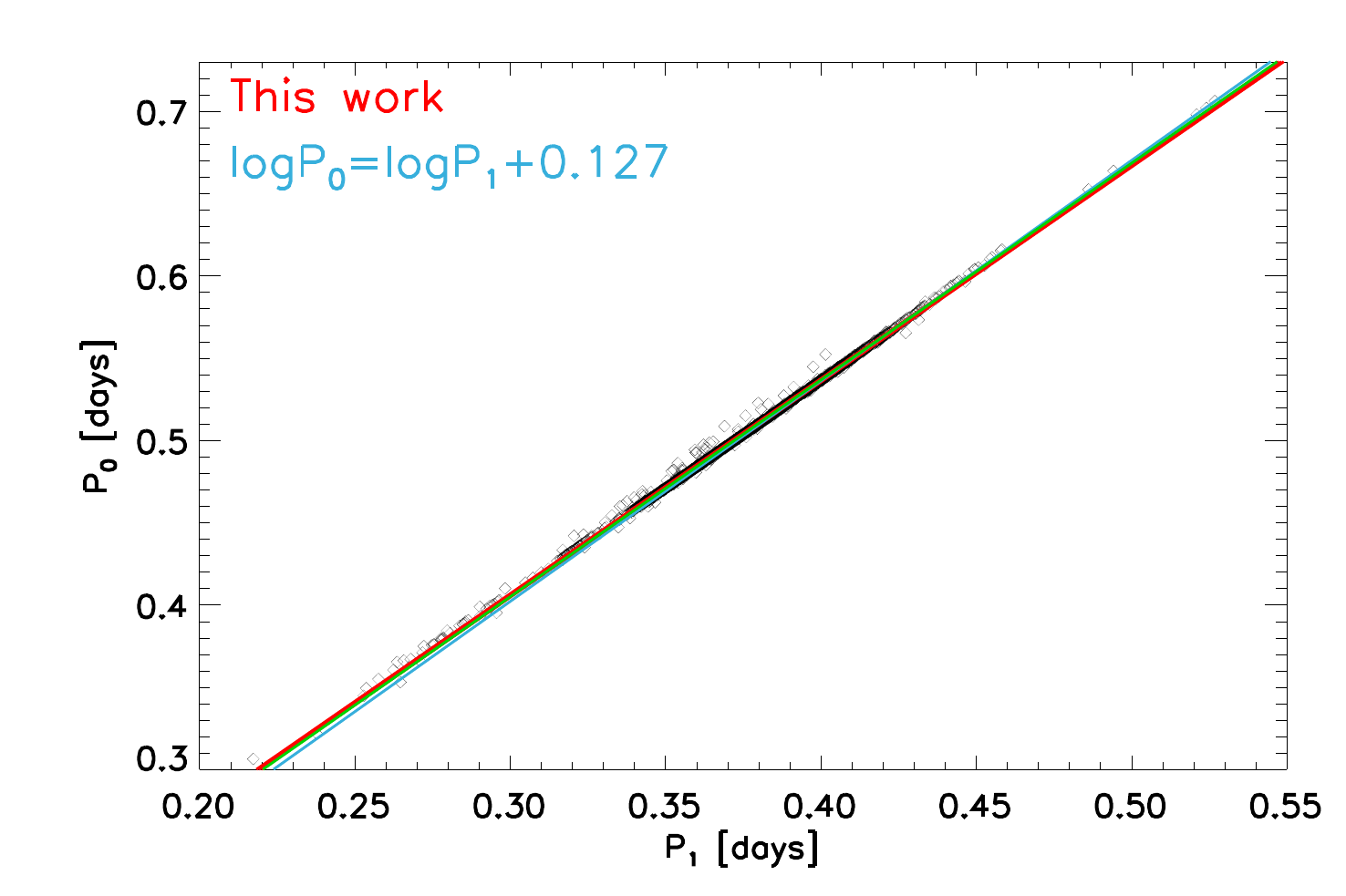}
    \caption{$P_0$ versus $P_1$ for all the RRds in the dSphs, MCs and the IG. They are displayed as black diamonds. No color coding was adopted for the points because they overlap too much and the different colors would not be discernible. The fundamentalization relations from our fit of the data, and from the classical formula are displayed as red, and light blue solid lines, respectively.}
    \label{fig:fundamentalization}
\end{figure}

\subsubsection{New quantitative relations}

We exploit the large sample of RRd in nearby stellar systems to derive a new empirical relation to fundamentalize the periods of FO RRLs. The classical relation ($\log{P_0}=\log{P_1} + 0.127$), displayed in blue in Fig.~\ref{fig:fundamentalization}, dates back to more than 40 years ago \citep[][]{sandage81c,cox1983,petersen1991}, when it was noted that FU and FO periods display a constant ratio in RRds with different iron abundances. The logarithmic relation was adopted, in the first place, due to the very limited sample of RRds available and a simple constant ratio between periods (meaning a constant offset in $\log{P}$ was noticed). However, the availability of a large RRd sample to derive a fundamentalization relation, recently allowed \citet[][]{coppola15} to provide a linear relation between $P_1$ and $P_0$, based on a sample of RRd similar to our own (i.e., including RRds in dSphs), despite being smaller.

One can adopt the fundamentalization relation to transform the period of RRc variables and treat them as equivalent to RRab variables. This is a typical trick to increase the sample of RRLs used to calibrate the Period-Luminosity relations and/or to estimate the distance of a stellar system \citep[][]{longmore1990,dallora04,braga15}.

Thanks to the size of our sample, we could derive, as did by \citet[][]{coppola15} a linear relation by applying a linear fit to our data and we found:


\begin{equation}\label{eq:fund_emp}
\begin{split}
    P_0 = (0.0135 \pm 0.0003) + (1.3067 \pm 0.0008)\cdot P_1 \\ (\sigma=0.0011\; days)
\end{split}
\end{equation}


The two relations (the classical and our own) overlap well with the data in Fig.~\ref{fig:fundamentalization} and between themselves. The coefficients are also similar to those by \citet[][]{coppola15}, when the intrinsic dispersion is taken into account. In passing we also note that the minimal dispersion showed by the current sample is further supporting that FU and FO periods are characterized by a similar dependence on the metal content \citep[][]{fabrizio2021}. Indeed, their difference is constant over a broad range in pulsation periods and in metal abundances. 

\begin{figure}
	\includegraphics[width=\columnwidth]{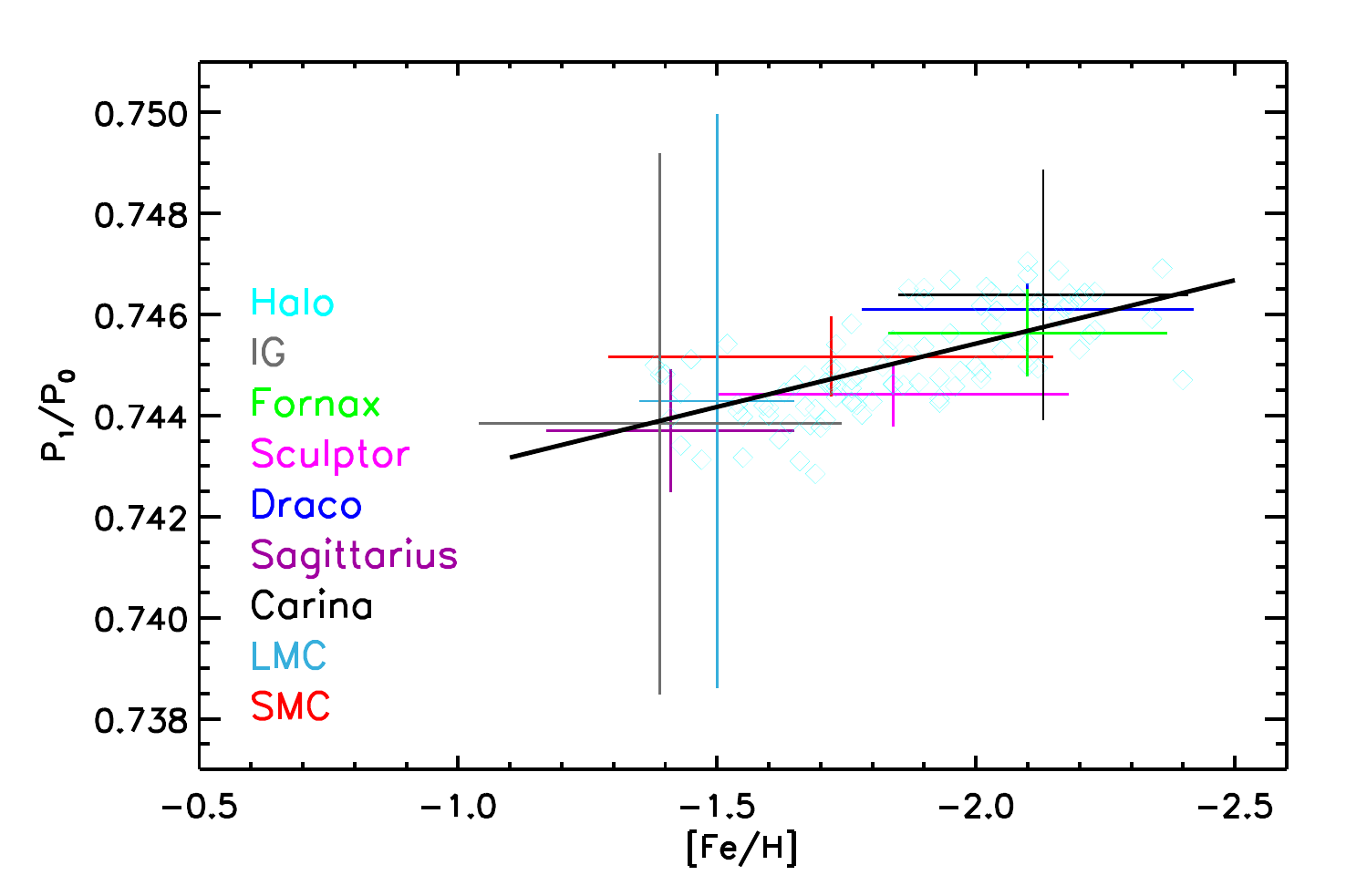}
    \caption{Period ratio versus metallicity for the dwarf galaxies in our sample, the IG and 95 Halo RRds. The galaxies and the IG are marked with large crosses, centered at the median $\dfrac{P_1}{P_0}$ ratio, with vertical and horizontal bars showing the $\sigma$. For the Halo RRds, we have homogeneous, individual iron abundance measurements \citep[1 from high-resolution spectroscopy and 94 from the $\Delta$S method][]{liu2020,crestani2021a,fabrizio2021}. The color coding is the same as in Fig.~\ref{fig:petersen} except for Halo stars that are displayed with cyan diamonds.}
    \label{fig:feh_vs_ratio}
\end{figure}

However, the period ratio $\dfrac{P_1}{P_0}$ does depend on the metallicity. Data plotted in Fig.~\ref{fig:feh_vs_ratio} show that the period ratio decreases when moving from metal-poor to metal-rich RRds. Therefore, we also investigate its dependence on the metallicity by using the iron abundances listed in Table~\ref{tab:dsph}.

For this purpose, we adopted the median ratios of the IG and dwarf galaxies in our sample, together with the iron abundances in Table~\ref{tab:dsph} and the individual ratios for 95 Halo RRds. We found not only an overlap between the individual Halo stars and the medians of the other samples, within the errors, but also a well defined linear trend. We performed a linear fit and we found

\begin{equation}\label{eq:ratio_feh}
\begin{split}
    \dfrac{P_1}{P_0} = (0.74041 \pm 0.00071) - (0.00251 \pm 0.00038)\cdot [Fe/H] \\ (\sigma=0.00072)
\end{split}
\end{equation}

 Note that the uncertainty on the slope of Eq.~\ref{eq:ratio_feh} is $\sim$15\% and the intrinsic spread of the period ratio is, at fixed iron abundance, similar to the spread in period ratio of the PD.

\subsection{Period-Amplitude RatioS diagram}\label{sec:specialdiagram}

The number of available phase points for Sculptor and Fornax allowed us to derive the $B$- and $V$-band amplitudes of the secondary modes ($Amp(B)_{sec}$ and $Amp(V)_{sec}$) of the RRds. Note that we are focusing our attention  on the canonical RRds, therefore, we use the subscript ``0'' instead of ``sec'' for the secondary and ``1'' instead of ``dom'' for the dominant mode. $Amp(B)_{0}$ and $Amp(V)_{0}$ were obtained by subtracting the model fit of the dominant mode, folding the residuals at the period of the secondary mode and fitting the folded light curve of the residuals. This means that we can estimate the amplitude ratio of the two modes as $Q(Amp(X))=\dfrac{Amp(X)_{1}}{Amp(X)_{0}}$, where $X$ indicates a generic pass-band.

\begin{figure*}
	\includegraphics[width=18.5cm,height=12cm]{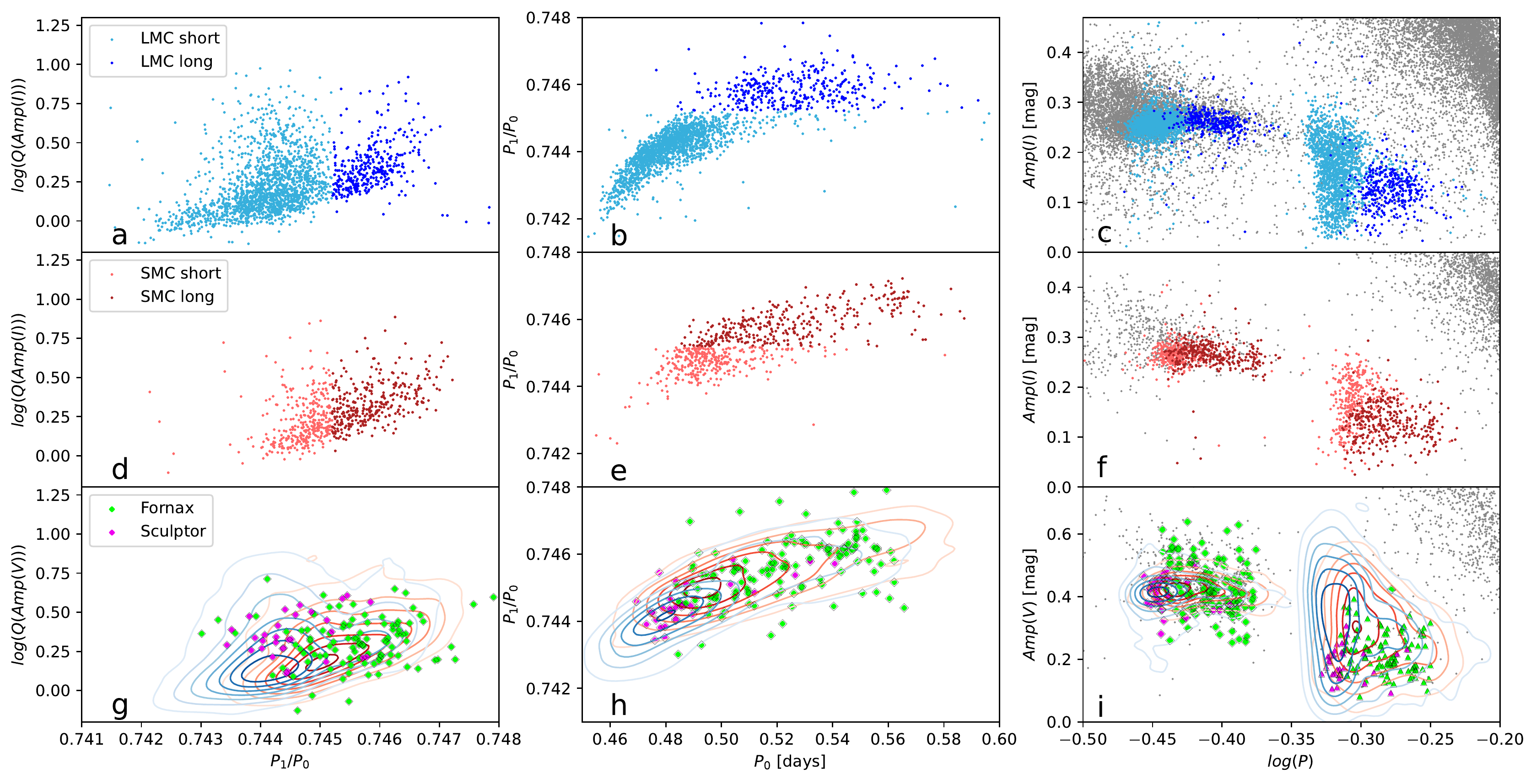}
    \caption{Left panels (a, d, g): PARS diagrams (amplitude ratio vs period ratio) of LMC (top), SMC (middle), Fornax and Sculptor (bottom) RRds. The light and dark symbols mark the positions of the \textit{short-} and \textit{long-period} (SP, LP) subgroups. The density contours of LMC (blue) and SMC (red) RRds are displayed for comparison in the bottom panel.
    Central panels (b, e, h): Petersen diagrams for LMC (top), SMC (middle) and Fornax plus Sculptor (bottom) RRds. The density contours of LMC (blue) and SMC (red) RRds are displayed for comparison in the bottom panel.
    Right panels (c, f, i): Bailey diagrams, luminosity amplitude versus logarithmic period, for LMC (top), SMC (middle) and Fornax plus Sculptor (bottom) RRds. The grey dots mark the position of RRc (top left) and RRab (top right) variables of the quoted galaxies. The luminosity amplitudes for both dominant and secondary component of \textit{short-} and \textit{long-period} RRds are plotted with the same colors and the same symbols adopted in the left panels. The density contours of LMC (blue) and SMC (red) RRds are displayed for comparison in the bottom panel. For Sculptor and Fornax RRds we adopted the $V$-band amplitudes (see \S~\ref{sec:fornax}). The triangles mark position, in the Bailey diagram, of the secondary period and amplitudes.}
    \label{fig:pars}
\end{figure*}

We show the Sculptor, Fornax, MCs and IG RRds in this new Period-Amplitude RatioS diagram (hereinafter PARS diagram), that is $log (Q(Amp(X)))$\footnote{We use the logarithm of the amplitude ratio for a more compact visualization.} versus the period ratio ($P_1/P_0$), in the left panels of Fig.~\ref{fig:pars}. 
To properly compare the PARS diagrams of different galaxies, with data collected in different bands, we have first checked, based on the 95 OGLE RRds with the highest number of $V$-band phase points, whether $Q(Amp(V))$ is equivalent to $Q(Amp(I))$. We first derived $Amp(V)_1$ and $/Amp(V)_0$ which are not provided within the OGLE catalogs, and we found a linear relation between the $V$- and $I$-band amplitude ratios: $Q(Amp(V)) = 0.14\pm0.05 + 1.02\pm0.02 \cdot Q(Amp(I))$, with an intrinsic dispersion of $\sigma=0.20$. This relation, within the standard deviation and the intrinsic errors, is consistent with $Q(Amp(V))\sim Q(Amp(I))$. Therefore, the ratios can be re-scaled and overlapped as in panel g) of Fig.~\ref{fig:pars}.

\subsubsection{The long- and short-period dichotomy}

A very interesting feature of the PARS diagram for LMC RRds is the bifurcation at $P_1/P_0\gtsim0.745$. The current data display that amplitude ratios are distributed along two sequences bending towards higher $Q(Amp(I))$: the former sub-group is located at  $P_1/P_0\sim0.744-0.745$ and the latter one at $P_1/P_0\sim0.746-0.747$. We name them ``short-period'' (SP) and ``long-period'' (LP) sub-group, according to their position in the PARS (panel a) of Fig.~\ref{fig:pars}). SMC RRd display a similar behaviour but the separation is less clear than for LMC RRd.
%

\begin{figure}
	\includegraphics[width=9cm]{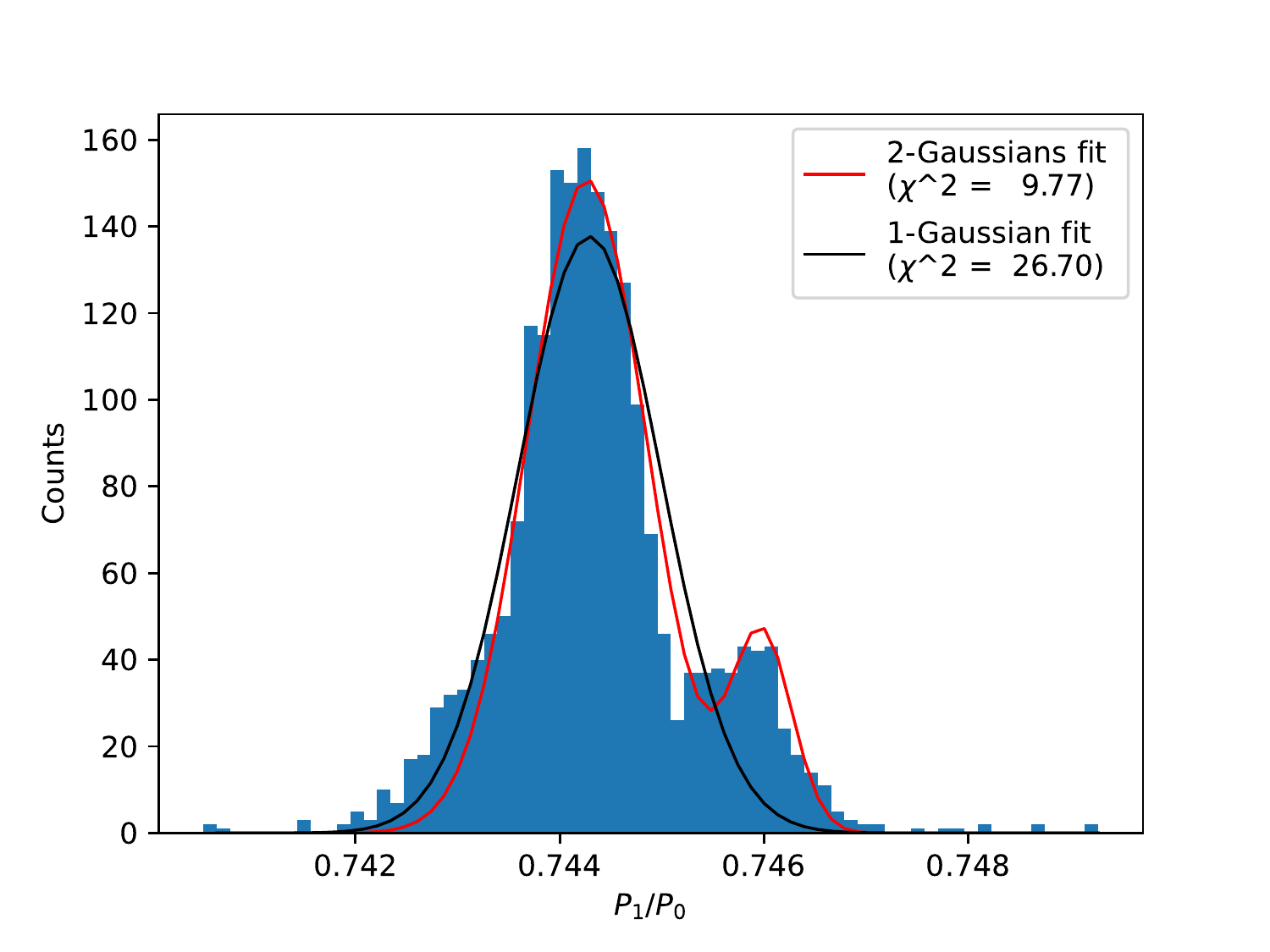}
    \caption{Histogram of the $P_1/P_0$ ratio of the RRds of Fornax. The black and red lines represent the fits with, respectively, one and two Gaussians. The $\chi^2$ of the fits are labelled.}
    \label{fig:gap}
\end{figure}

The separation between SP and LP is even more interesting when the two groups of variables are  plotted in the PD (panels b, e and h in Fig.~\ref{fig:pars}). Indeed, the LMC RRds belonging to the SP and the LP sub-groups appear to be distributed in two different regions with a clear separation at $P_1/P_0\sim0.745$. 
To investigate the bifurcation of LMC RRds in a more quantitative way, we analyzed the period ratio---$P_1/P_0$---distribution (see Figure~\ref{fig:gap}). The period ratio distribution was fit  with a single Gaussian (black line) and with two Gaussians (red line). The latter performs a proper fit of both primary (SP) and secondary (LP) peak---centered, respectively, at 0.7443 and  0.7460---and the $\chi^2$ decreases by a factor of more than two. This finding soundly supports the dichotomic distribution between SP and LP RRds.

The separation is once again less evident for SMC RRds. The distribution is quite homogeneous over the entire period range, moreover, the region located at $P_1/P_0\sim0.745$ and $P_0\sim0.495$ days is well populated in the SMC, while in the LMC is only populated by a few variables. There is no firm evidence of a dichotomy in the metallicities of the old stars in the LMC. \citet[][Fig.17]{gratton04a} show different metallicity distributions for RRcs and RRds (two-peaks distribution) with respect to RRabs (asymmetric one-peak distribution) in the LMC. However, their statistics are too low (RRab=64, RRc=27, RRd=7) to draw firm conclusions.
Thus, we can explain the dichotomy observed in the LMC PARS diagram as a difference in metallicity distribution given the well-known ranking with metallicity in the PD \citep{coppola15} and the existence of a period-metallicity relation observed in  RRLs \citep[][]{arp1955,preston1959,fabrizio2021}. 
The population ratio (i.e., the ratio between the number of SP and LP variables, $n^{SP}_{LP}$) is $\sim$4.29 for the LMC and $\sim$1.04 for the SMC. This suggests that the metal-rich (SP) component among the LMC RRds is more prominent than the metal-poor (LP) one, while for SMC RRds the metal-rich and the metal-poor components attain similar values.
The SP and the LP sub-groups have also different properties in the Bailey diagram (panels c,f,i in Fig.~\ref{fig:pars}). The $Amp(I)_1$ attains similar values for the two sub-groups while the $Amp(I)_0$ cover a broader range. This trend is quite clear for LMC RRds, where the $Amp(I)_0$ range covered by SP RRds is a factor of two larger than for LPs.  Note that a broad variation in metal content inside the SP sub-group can be ruled out, because RRc luminosity amplitudes are almost a factor of two more  sensitive to [Fe/H] than  RRab variables \citep[][]{fabrizio2021}. Therefore, a large spread in metal content should show up with a spread in $Amp(I)_1$ larger than in $Amp(I)_0$. 

As a working hypothesis, the observed spread in $Amp(I)_0$ seems to be connected with a difference in the topology of the so-called OR region when moving from more metal-poor to more metal-rich RRds \citep[][]{bono97d}. Indeed, the current evidence seems to indicate that the SP sub-group is approaching a mode change, when compared with the LP sub-group. This would also imply that the period derivatives of the former sample should be on average larger than the latter one.

Finally, let us mention that the quoted difference cannot be explained by photometric errors, because the $Amp(I)_0$ of the LP sub-group are, on average, smaller and display a smaller dispersion. We also checked whether $Amp(I)_0$ could depend on either $P_1/P_0$, or on de-reddened mean $(V-I)_0$ color (by using reddening maps by \citealt[][]{skowron2021}) or any other pulsation parameter available, but they do not display any evident correlation. The increased spread in  $Amp(I)_0$ amplitudes is also present in SMC RRds, but less evident due to limited statistics. 

The current evidence suggests that the PARS diagram could be adopted to probe the metallicity distribution of old stellar populations producing RRLs. However, the PARS diagrams for LMC RRds (left panels of Fig.~\ref{fig:pars}) bring forward new features that are worth to be discussed in more detail. The SP subgroup shows, at fixed period, a larger spread in amplitude ratios when compared with the LP subgroup. However, the comparison between theory and observations showed in Fig.~\ref{fig:petersen} suggests that the variation in chemical composition within the same subgroup is quite limited. This indicates that the main culprit for the observed spread might be the surface gravity. HB evolutionary models show that the surface gravity of low-mass stellar structures changes during their off-ZAHB evolution \citet{bono2020b}. Recent results  based on the Bailey diagrams taking account for optical, mid-infrared and radial velocity amplitudes, suggest that part of the spread, at fixed pulsation period, is caused by evolutionary effects \citep[][]{bono2020}. The PARS diagram is providing a new and more detailed view on this long-standing problem. The surface gravity linearly scales with $\mathcal{M}$, but it is inversely proportional to $R^2$. The pulsation period, according to the Ritter relation \citep[][]{ritter1879}, scales as $R^{3/2}$, meaning that evolutionary effects would imply a change both in amplitude ratio and in pulsation period. Moreover and even more importantly, the current empirical and theoretical evidence indicate that RRds are distributed within a narrow and well-defined sequence in the PD. Thus suggesting that the variation in period is mainly caused by evolutionary effects driven by variations in chemical composition. 

We are left with the circumstantial evidence that the spread in the PARS diagram seems mainly caused by variation in surface gravity. Indeed, the same variables display a modest spread in the Petersen diagram and a larger spread in the PARS diagram, thus suggesting that the amplitude ratios are a diagnostic more sensitive to surface gravity.

\subsubsection{Fornax and Sculptor}

We note that, in the PARS diagram, the RRds of Fornax and Sculptor overlap with the LP and the SP sub-group observed in the LMC, but do not show any clear separation. Note that metallicities listed in Table~\ref{tab:dsph} suggest that Sculptor and Fornax cover the metal-poor tail of LMC and SMC RRds. We also point out that Sculptor RRds in the PARS diagram do not perfectly overlap with the highest density region of LMC RRds, having a larger $Q(Amp(V))$.

The density contours of the LMC and SMC in the PD (panel h) outline an interesting feature that was not immediately clear in Figure~\ref{fig:petersen}. The distribution of Sculptor and LMC RRds is similar: both have their peak in density around $P_0 \sim 0.48$ days and they are extended to longer periods, but with lower density. The main difference is that the short-period tail of the LMC ($P_0 \lesssim 0.46$ days) is still highly populated, while the distribution of Sculptor RRds seems to be truncated. Although, the number of RRds in these two galaxies is quite different (30 vs 2083 objects) one can safely assume that the difference is not just a matter of statistics. There are reasons to believe that the difference in total mass, and in turn, in its chemical enrichment (mass-metallicity relation), between the two galaxies \citep[][]{mcconnachie12} is preventing the production of more metal-rich RRds in Fornax. Nonetheless, the two distributions in the PD are similar. The same outcome applies to Fornax RRds, but they closely resemble the distribution in the Petersen Diagram of SMC RRds (see red contours).

The similarities discussed in the middle panels of Fig.~\ref{fig:pars} also show up in the Bailey diagram plotted in the right panels of the same figure. Fornax RRds display a large dispersion in $Amp(V)$, in both modes when compared with Sculptor RRds. Note also that the distribution of FU amplitudes in Sculptor RRds is not as extended as that of the LMC RRds and they cluster at lower $Amp(V)_1$. This explains why, as already mentioned, Sculptor RRds have larger $Q(Amp(V))$ with respect to the LMC. As for the PD, it is not possible to draw firm conclusions, due to the difference in number of objects and in sampling (OGLE light curves have at least one order of magnitude more phase points than ours). In passing, it is worth mentioning that, for a proper comparison, the contours in panel i) were rescaled by a factor of 1.59, that is the typical value for $Amp(V)$/$Amp(I)$ \citep[][]{braga16}.

\section{Conclusions}

We have derived the pulsation properties of the RRds in Fornax and in Sculptor dSph and complemented these data with literature properties of RRds in other dSphs, in the MCs, in the Halo and in the IG. This allowed us to inspect the behavior of RRds in the PD, in the Bailey diagram and in the brand new PARS diagram.

{\it Fornax and Sculptor RRds---} Based on $BV$-band photometric data collected during more than 20 years we have, for the first time, identified and characterized Fornax RRds. We have found 130 canonical and 212 candidate RRds among 2068 RRLs. We also provided an updated catalog of 32 canonical RRds in Sculptor including their pulsation properties, based on the photometry by \citep[][]{martinezvazquez15}.

{\it Metallicity trend of $\eta_{RRd}$---} The population ratio---the fraction of RRds over the total number of RRLs---in metal-poor stellar systems is on average larger than for metal-rich stellar systems. This finding supports the trend with the metallicity originally suggested by \citet[][]{coppola15}. The statistics concerning GGCs is too limited to rich firm conclusions concerning the dependence of the population ratio on the metallicity.


{\it Fornax RRds fill the gap of dSphs---} The RRds of Fornax span a large range in $P_0$ (between 0.49 and 0.57 days), in which they are uniformly distributed. This suggests that Fornax RRds have a metallicity distribution that is more uniform than other dSphs. Moreover, Fornax RRds fill the gap previously found for dSph RRds at $0.49 d<P_0<0.53 d$. One Carina RRd and a couple of Sagittarius RRds fall within the gap, however these should be considered as isolated objects.  By comparing the distribution of dSph and Halo RRds in the PD, we find that they are more alike without the gap and this agrees with previous results obtained by \citet{fiorentino2017b}, suggesting that less massive dSphs contributed at most $\sim$20\% of the Halo mass assembly.

{\it Fundamentalization relation---} Based on 3396 RRds---that is a catalog more than twice as large as in previous works---including all the RRds in the dSphs, MCs and the IG, we obtained a linear relation of  that overlaps with the canonical, logarithmic one. The uncertainties on the coefficients and the $\sigma$ of our relation are smaller (from a 30\% to a factor of ten) than those of the theoretical relation. We quantitatively show that this relation is not affected by metallicity, while the ratio $\dfrac{P_1}{P_0}$ does show a clear anticorrelation with iron abundance.

{\it RRd dichothomy---} By inspecting the new PARS diagram, where the ratio of FO over FU amplitudes in a given pass-band $X$  ($Q(Amp(X))$) is plotted versus the period ratio ($P_1/P_0$), we found a clear dichotomy in the RRds in the LMC. We name the two groups Short-Period (SP) and Long-Period (LP) due to their position in the PD. As a working hypothesis we suggest that variations in surface gravity, caused by evolutionary effects, causes the spread in amplitude ratios observed in the PARS diagram. More quantitative constraints require accurate dynamical (mass) and interferometric measurements. LMC: The two groups are well-separated at $P_1/P_0$=0.745 and their number ratio is $n^{SP}_{LP} \sim 4.29$. The PD shows a clear separation between the LP and the SP subgroup, thus suggesting that this is a dichotomy and not a smooth transition. In the Bailey diagram, the SPs and LPs are also clearly separated. Moreover, the $Amp(I)_0$ of SPs cover a wider range than those of the LPs, while $Amp(I)_1$ of LPs and SPs are similar. We suggest that this difference is caused by a difference in the topology of the "OR region" when moving from the more metal-poor to the more metal-rich regime. SMC: The transition between SPs and LPs does not show a clear separation and the number ratio---as expected by the lower metallicity of the SMC---is lower ($n^{SP}_{LP} \sim 1.04$). Fornax and Sculptor: The RRds of these two dSphs overlap, in all plots, with the SMC and LMC, respectively, in qualitative agreement with their metallicities. 

The PARS diagram, together with the Petersen and the Bailey diagrams, are powerful distance- and reddening-independent diagnostics to discriminate physical properties of old stellar populations  in resolved stellar systems. The coupling between the PARS and the Petersen diagrams is a very powerful diagnostic to investigate old stellar populations in stellar systems hosting a sizable sample of RRds. The current evidence indicates that LMC RRds are split into different subgroups characterized by two different metal contents. This indicates that old stellar populations in the LMC were formed in two consecutive, but different star formation episodes. Moreover and even more importantly, the latter event was significantly more relevant concerning the the fraction of mass involved. Indeed, the SP (more metal-rich) sample is four times larger than the more metal-poor one. On the other hand, the SMC RRds in the SP and in the LP subgroup are very similar, thus suggesting that they also formed in two consecutive star formation events, but the mass involved was quite similar. To our knowledge this is the first time that  a pulsation diagnostic can be used to trace back in time and in chemical composition the properties of old, low-mass stellar populations. Accurate and homogeneous spectroscopic measurements can shed new lights on this interesting new path.

Finally let us mention that the opportunity to fully exploit RRds as old stellar tracers appears even more appealing when focusing on the Halo and it could not have been more timely than this since Gaia full DR3---including new photometric time series---was recently released \citep[][]{gaia_dr3}, and the Vera C. Rubin Observatory Legacy Survey of Space and Time (LSST) will begin soon.

\section*{Acknowledgements}

V.F. Braga acknowledges the financial support of the Istituto Nazionale di Astrofisica (INAF), Osservatorio Astronomico di Roma, and Agenzia Spaziale Italiana (ASI) under contract to INAF: ASI 2014-049-R.0 dedicated to SSDC.
V.F. Braga and M. Monelli acknowledge financial support from the ACIISI, Consejer\'ia de Econom\'ia, Conocimiento y Empleo del Gobierno de Canarias and the European Regional Development Fund (ERDF) under the grant with reference ProID2021010075.
M. Marengo was partially supported by the National Science Foundation under grant No. AST1714534.
C.E. Mart{\'i}nez-V{\'a}zquez is supported by the international Gemini Observatory, a program of NSF's NOIRLab, which is managed by the Association of Universities for Research in Astronomy (AURA) under a cooperative agreement with the National Science Foundation, on behalf of the Gemini partnership of Argentina, Brazil, Canada, Chile, the Republic of Korea, and the United States of America.

\section*{Data Availability}

The data underlying this article are available in the article and in its online supplementary material.



\bibliographystyle{mnras}
\bibliography{ms} 





\bsp	
\label{lastpage}
\end{document}